\documentclass{aa}

\usepackage{graphicx}
\usepackage{subcaption}
\usepackage[dvipsnames]{xcolor} 
\usepackage{txfonts}
\usepackage[colorlinks=true
  ,urlcolor=blue
  ,anchorcolor=blue
  ,citecolor=blue
  ,filecolor=blue
  ,linkcolor=blue
  ,menucolor=blue
  ,linktocpage=true
  ,pdfproducer=medialab
  ,pdfa=true
]{hyperref}

\usepackage{multirow}
\usepackage{stfloats}
\usepackage{adjustbox}  

\usepackage[normalem]{ulem}

\definecolor{tangerine}{rgb}{1.0,0.57,0.0}

\begin{document}

   \title{Accurate $N$-body simulations with local primordial non-Gaussianities: Initial conditions and aliasing}

   \author{Adrian Gutierrez Adame \inst{1,2,3}
          \and 
          Santiago Avila\inst{4,5}
          \and
           Violeta Gonzalez-Perez\inst{1,2}
          \and
          Oliver Hahn \inst{6,7}
          \and 
          Gustavo Yepes \inst{1,2}
          \and 
          Marc Manera \inst{4,8}
          }

   \institute{
            Departamento de F\'isica Te\'orica,  Universidad Aut\'onoma de Madrid, 28049 Madrid, Spain
         \and
            Centro de Investigaci\'on Avanzada en F\'isica Fundamental (CIAFF), Facultad de Ciencias, Universidad Aut\'onoma de Madrid, ES-28049 Madrid, Spain
            \and 
            Instituto de F\'isica Teorica UAM-CSIC, c/ Nicolás Cabrera 13-15, Cantoblanco, 28049 Madrid, Spain
             \and
            Institut de Física d’Altes Energies (IFAE) The Barcelona Institute of Science and Technology campus UAB, 08193 Bellaterra Barcelona, Spain
        \and 
            Centro de Investigaciones Energ\'eticas, Medioambientales y Tecnol\'ogicas (CIEMAT), Madrid, Spain
        \and 
            Department of Astrophysics, University of Vienna, Türkenschanzstrasse 17, 1180 Vienna, Austria 
        \and 
            Department of Mathematics, University of Vienna, Oskar-Morgenstern-Platz 1, 1090 Vienna, Austria               
        \and 
            Serra H\'unter Fellow, Departament de F\'isica, Universitat Aut\'onoma de Barcelona, Edifici C Facultat de Ciències, 08193 Bellaterra, Spain          
            }
   \date{Submitted June 6, 2025}
    \titlerunning{Accurate $N$-body simulations with local Primordial non-Gaussianities: initial conditions and aliasing}
    \authorrunning{A.G. Adame et al.}

\abstract{The new generation of galaxy surveys designed to constrain local primordial non-Gaussianity (PNG) requires $N$-body simulations that accurately reproduce its effects. In this work, we explore various prescriptions for the initial conditions of simulations with PNG, seeking to optimise accuracy and minimise numerical errors, particularly due to aliasing. 
We used $186$ runs that vary the starting redshift, Lagrangian Perturbation Theory (LPT) order, and non-Gaussianities ($f^{\rm local}_{\rm NL}$ and $g^{\rm local}_{\rm NL}$). Starting with third order LPT ($3$LPT) at a redshift as low as $z_{\rm start}\simeq 11.5$ reproduces to $<1 \%$ the power spectrum, bispectrum, and halo mass function of a high‑resolution reference simulation. The aliasing induced by the PNG terms in the power spectrum produces a $ \leq 3 \%$ excess for small‑scales at the initial conditions. This excess drops below $0.1\%$ by $z=0$. State-of-the-art initial condition generators show a sub-percent agreement. 
We show that initial conditions for simulations with PNG should be established at a lower redshift using higher-order LPT schemes. We also show that removing the PNG aliasing signal is unnecessary for current simulations.
The methodology proposed here can accelerate the generation of simulations with PNG while enhancing their accuracy.}
  

   \maketitle

\section{Introduction}

Numerical simulations are a powerful tool for cosmology, as they can accurately model structure formation deep into the non-linear regime ($k> 1 \,h\,\mathrm{Mpc}^{-1}$), where perturbative analytical methods break down. Today, simulations are fundamental at nearly every phase of large-scale structure (LSS) analysis. They play a key role in developing data processing pipelines for extracting the cosmological information \citep[e.g.][]{Noriega_2024,Krause_2017,Alam_2021,Riquelme_2022}, estimating the covariance matrices \citep{Avila_2018,Zhao_2021,Ereza_2024,Forero_2025}, and evaluating the impact of observational systematics \citep{Ross_2017,Bianchi_2018,Chan_2022,Chaussidon_2024}.

Until recently, the statistical power of galaxy surveys was the main bottleneck for cosmological inference \citep{Abell_2009,Weinberg_2013}. However, the current limiting factor is shifting towards theoretical and numerical modelling, as seen in experiments such as the Dark Energy Spectroscopic Instrument (DESI) \citep{Aghamousa_2016} and Euclid \citep{Laureijs_2011}, as well as in forthcoming facilities such as the Rubin Observatory  \citep{Abell_2009}. Improving analytic models demands parallel advances in simulation accuracy, where a good understanding of the parameters that control numerical precision is required \citep{Power_2003,Heitmann_2010,Schneider_2016}.

Primordial non-Gaussianity (PNG) is a key target of upcoming surveys, offering a direct window into inflationary physics. The simplest single-field slow-roll models for inflation predict that curvature perturbations are nearly Gaussian \citep{Baumann_2012,Riotto_2017}. This implies that all the statistical information is contained in the two-point correlation function or its Fourier equivalent, the power spectrum. One way to look for departures from Gaussianity is to study the three-point function (bispectrum). In standard single-field scenarios, the bispectrum amplitude, $f_{\rm NL}$, in the squeezed limit $(k_1\sim k_2 \gg k_3)$ is $\mathcal{O}(n_{s} -1)$ \citep{Maldacena_2003,Creminelli_2004}. Thus, detecting $|f_{\rm NL}| \gtrsim 1$ would be enough to rule out canonical single-field slow-roll inflation. At the same time, a measured value of $f_{\mathrm{NL}} \sim \mathcal{O}(n_{s} -1)$ would severely constrain other inflationary models \citep{Bartolo_2004,Byrnes_2010b}.

Models that propose that inflation is driven by multiple fields can naturally lead to $|f_{\rm NL}| \sim 1$ \citep{Byrnes_2010,Chen_2010}. An example of these models is the curvaton inflation \citep[e.g.][]{Lyth_2002,Lyth_2003}. Such scenarios can be parametrised in terms of the local PNG as
\begin{equation}
\phi_{NG}(x) = \phi_{G}(x) + f^{local}_{\rm NL} \left(\phi_{G}^2(x) + \langle\phi_G^2\rangle\right),
\end{equation}
where $\phi_G$ is the Gaussian primordial gravitational potential and $f_{\rm NL}^{local}$ controls the non-Gaussian quadratic correction \citep{Komatsu_2001,Salopek_1990}. 

Currently, the strongest bounds on local-type PNG come from measurements of the bispectrum of temperature fluctuations of the cosmic microwave background (CMB), leading to $f_{\rm NL} = -0.9 \pm 5.1\; (68\%\,{\rm c.l.)}$ \citep{Planck_2018b}. However, the amount of information that we can extract from the CMB about local PNG is limited by cosmic variance. For this reason, it is not expected to go beyond $\sigma(f_{\rm NL})\sim 3$ with CMB-only data \citep{Komatsu_2001,Babich_2004}. An alternative method has been proposed to beat these constraints by analysing the LSS of the Universe. The core idea is to use the scale-dependent bias ($b^2 = P_{gg}(k)/P_{mm}(k) \propto 1/k^{4}$ at large scales) that local PNG induces in galaxy clustering \citep{Dalal_2008,Slosar_2008,Matarrese_2008}. 

The most stringent constraints from the LSS method come from the first year of data of the DESI survey, where they found $f_{\rm NL} = -3.6^{+9.0}_{-9.1}\; (68\%\,{\rm c.l.})$ \citep{Chaussidon_2024}. Forecasts for Stage IV surveys including information from the bispectrum \citep{Tellarini_2016,Karagiannis_2018}, multi-tracer techniques \citep{Barreira_2023,Sullivan_2023,Karagiannis_2024}, and data from upcoming experiments such as SphereX \citep{Dore_2015,Shiveshwarkar_2024} are expected to reach $\sigma(f_{\rm NL})\sim 0.5$ \citep{Ballardini_2019,Heinrich_2024}.

To distinguish between the more canonical inflationary models and other scenarios, new surveys have been designed with increasingly smaller target $\sigma(f_\mathrm{NL})$, for example SphereX \citep{Dore_2015} and SPEC-S5 \citep{Besuner_2025}. As a result, simulations that accurately include PNG are growing in importance. Significant efforts have been made to produce realistic simulations for these surveys \citep[e.g.][]{Adame_2024,Hadzhiyska_2024,Bayer_2024}. A key challenge in running simulations in general, and PNG simulations in particular, is ensuring they are accurate enough. In the past, the community carried out large code comparison projects by running the same initial conditions with different $N$-body codes, leading to a $1\%$ agreement among them in the matter power spectrum at $k=10 \, h\,{\rm Mpc}^{-1}$ \citep[e.g.][]{Schneider_2016,Garrison_2019,Springel_2021}. 

Building on these convergence benchmarks, attention has now shifted to mitigating the numerical artefacts that remain a threat to sub-percent accuracy in forthcoming PNG simulations. Discreteness effects from the particle lattice self-interactions have been a prominent concern \citep{Joyce_2005,Joyce_2007,Garrison_2016}. Several proposals have been made to overcome these issues with tessellation methods \citep{Hahn_2013,Hahn_2016b,Sousbie_2016,Stucker_2020}. $N$-body codes can be affected by significant numerical errors when computing forces against  nearly uniform density fields at a high redshift. This motivates the idea of pushing the initial redshift to lower redshifts, where these effects are suppressed \citep{Schneider_2016,Michaux_2020,Stahl_2024}.

In this work we pursue three goals: (i) validating how the initial redshift and LPT order affect convergence in the presence of local PNG, (ii) quantifying PNG aliasing effects, and (iii) testing consistency among state‑of‑the‑art codes for IC generation.  The PNG initial condition generation step adds its own subtleties. The computation of the fields $\phi^2$ and $\phi^3$ on a discrete grid excite modes that cannot be represented in the original grid and fold back. This contaminates higher-frequency modes and induces a spurious signal. This effect is known as aliasing. The aliasing of the power spectrum and bispectrum estimation , and the computation of higher-order corrections in the LPT expansion has been well studied in the past \citep[e.g.][]{Sefusatti_2016, Michaux_2020}. However, the majority of previous studies using PNG simulations have ignored the aliasing due to PNG altogether. Although some studies have accounted for (and corrected for) the aliasing \citep[e.g.][]{Stahl_2023,Stahl_2023b,Stahl_2024b,Stahl_2025}, to our knowledge its impact has not been explicitly quantified before. We also compared different state-of-the-art codes for generating the initial conditions for PNG simulations, namely \textsc{MonofonIC} \citep{Michaux_2020,Hahn_2020}, \textsc{2LPTPNG} \citep{Crocce_2006,Scoccimarro_2012}, \textsc{FastPM} \citep{Feng_2016},  and \textsc{LPICOLA} \citep{Howlett_2015}.

This paper is structured as follows. In \autoref{sec:methods} we introduce how initial conditions are generated for cosmological simulations. We also discuss how this is done for PNG simulations, particularly how aliasing arises in these cases, and how it can be corrected. We then describe in \autoref{sec:simulations} the simulation suite that we have produced to demonstrate all these points. We present our results in \autoref{sec:results}. We first discuss the impact of the initial redshift and LPT order. Then, we demonstrate the impact of aliasing, and compare some of the state-of-the-art codes for initial condition generation. Finally, we summarise our results and draw our conclusions in \autoref{sec:conclusions}.

\section{Methods}\label{sec:methods}

Cosmological $N$-body simulations are essential for tracking the growth of structures deep into the non-linear regime, where perturbative techniques fail. State-of-the-art codes such as \textsc{PKDGRAV3} \citep{Potter_2017} and \textsc{Gadget4} \citep{Springel_2021} integrate the equations of motion for a large number of phase-space tracers (particles), capturing the evolution from the smooth early density field to the web of filaments, sheets, and virialised halos observed today.

To generate the initial conditions (ICs) for these simulations, we must specify the positions and velocities of every particle at a given point in the past. The accuracy of these ICs determines the upper limit on the precision for any late-time statistic that can be modelled with the simulation. Typically, a random realisation of the density field matching the linear power spectrum is generated, which then is back scaled to the desired initial redshift. Then the displacements and velocities of the particles are obtained using the Lagrangian perturbation theory truncated at some low order. In what follows, we describe these standard algorithms used to build Gaussian ICs and then show how they can be extended to include local-type primordial non-Gaussianity.

\subsection{Generating initial conditions for N-body simulations}

The most popular method to initialise cosmological $N$‑body simulations is the back‑scaling method \citep[see Sec. 6 from][for a review]{Angulo_2022}. The first step is to solve the coupled Einstein–Boltzmann equations using, for example, \textsc{CAMB} \citep{Lewis_2000} or \textsc{CLASS} \citep{Lesgourgues_2011}. These codes evolve all relevant species (cold dark matter, baryons, photons, massive neutrinos,~\dots) down to the target redshift for the final desired output of the simulation, $z_{\mathrm{target}}$. The resulting linear total‑matter density contrast $\delta_m(k,z_{\mathrm{target}})$ is then rescaled to the initial redshift $z_{\rm start}$ at which the $N$‑body run begins,
\begin{equation}
  \tilde{\delta}_m(k,z_{\rm start})=
  \frac{D_{+}(z_{\rm start})}{D_{+}(z_{\mathrm{target}})}\,
  \delta_m(k,z_{\mathrm{target}})\;,
\end{equation}
where $D_{+}(z)$ is the linear growth factor normalised to
$D_{+}(0)=1$. By construction, the back‑scaled field reproduces the exact linear solution at $z_{\mathrm{target}}$, including scale‑dependent effects of radiation and massive neutrinos, while neglecting the decaying modes. These modes are already strongly suppressed by $z_{\rm start}$ if starting late enough. Although this procedure yields a fictitious radiation content at the initial time, it remains fully consistent with relativistic perturbation theory when the system is evolved forward with Newtonian dynamics \citep{Chisari_2011,Hahn_2016,Fidler_2017}.

To generate the initial conditions, we have to specify a particular realisation of the density field $\delta_m$. This is modelled as a Gaussian random field (if there are no PNGs) whose variance is given by the power spectrum. We begin with a complex random white noise field, $W(\mathbf{k})$, given by

\begin{equation}
  W(\mathbf{k}) \curvearrowleft \mathcal{N}_{\mathbb{C}}\left(0,1\right),
\end{equation}
where both, real and imaginary parts are sampled from a Gaussian distribution with zero mean and unit variance. The power spectrum in the density field is induced via
\begin{equation}
  \delta_m(\mathbf{k}) = \sqrt{P_{\rm lin}(k)}\,W(\mathbf{k}),
  \end{equation}
where $P_{\rm lin}(k,z_{\mathrm{start}})$ is the desired linear matter power spectrum obtained from the Boltzmann solver. Then, to make sure that $\delta_m(\mathbf{x})$ is real, we impose $\delta_m(\mathbf{k}) = \bar{\delta}_m(\mathbf{-k})$. 

When we work with PNGs, it is often more convenient to operate with the primordial gravitational potential $\phi$ instead of the matter density contrast $\delta_m$. These two quantities can be used interchangeably as they are related via the Poisson equation. In Fourier space, this interchangeability can be expressed as
\begin{equation}
    \delta(\mathbf{k},z)=\alpha(k,z)\,\phi(\mathbf{k}),
    \label{eq:poisson_fourier}
\end{equation}
where $\alpha(k)$ is defined as
\begin{equation}
  \alpha(k,z)=
  \frac{2\,D_{+}(z)}{3\,\Omega_{m,0}}\,\frac{c^{2}}{H_{0}^{2}}\,\frac{g(0)}{g(z_{\mathrm{rad}})}\,
  k^{2}\,T(k).
\end{equation}

Here $H_0$ is the Hubble parameter, $\Omega_{m,0}$ is the matter density at $z=0$ and $c$ is the speed of light. $T(k)$ is the matter transfer function normalised to $T(k\to 0)=1$, $D_+$ is the growth factor normalised to $D(z=0)=1$, and $g(z)=(1+z)D_+(z)$. The term $\frac{g(z_{\mathrm{rad}})}{g(0)}$ arises precisely because of the choice of this normalisation for the growth factor. We evaluate $g(z)$ at  radiation domination, so for the cosmology chosen here (see \autoref{tab:cosmology}), the value of $\frac{g(z_{\mathrm{rad}})}{g(0)}$ is $\sim 1.2741$.

With these definitions, the linear matter power spectrum can be written compactly as
\begin{equation}
  P_{\rm lin}(k,z)=\alpha^{2}(k,z)\,P_{\phi}(k),
\end{equation}
where $P_\phi(k)$ is the power spectrum of the primordial gravitational potential. This is set during inflation via 
\begin{equation}
    \Delta_\phi^{\,2}(k)\equiv k^{3}P_\phi(k)/(2\pi^{2}) = A_s\,(k/k_\ast)^{\,n_s-1}
\end{equation}
with $k_\ast$ the pivot scale and $A_s$ the amplitude of the primordial fluctuations, such that $P_\phi \propto k^{n_s-4}$.

The stochasticity of the white noise is what leads to the cosmic variance in the simulations. One can take advantage of how this noise is generated to suppress much of this variance. For example, generating two realisations where phases are shifted by a factor of $\pi$ with respect to each other leads to inverted realisations (i.e. high-density peaks in one simulation become voids and vice versa) \citep{Pontzen_2016}. This is precisely what we do in this paper to improve the measurements of the summary statistics. 

Another option is to remove all the stochasticity in the amplitude part of the white noise, leading to a realisation with suppressed variance. Combined with the previous one, this idea composes the fixed-and-paired method \citep{Angulo_2016}.  Moreover, keeping the same white noise seed across simulations with different cosmologies (e.g. $f_{\mathrm{NL}}=0$ vs $f_{\mathrm{NL}}\neq0$) leads to correlations that can be used to improve the measurements of cosmological parameters via the difference between simulations \citep{Avila_2022,Adame_2024}.

Once the realisation of the initial overdensity field (or primordial potential) is generated, the next step is to obtain the initial displacements and velocities for the particles. For that,  the particles are first placed on an unperturbed cubic lattice, $\mathbf{q}$, representing the homogeneous Universe at $a\to 0$. Their Eulerian positions at $z_{\mathrm{start}}$ follow from Lagrangian perturbation theory (LPT),
\begin{equation}
  \mathbf{x}(\mathbf{q},a)=
  \mathbf{q}+
  \mathbf{\Psi}^{(1)}(\mathbf{q},a)+
  \mathbf{\Psi}^{(2)}(\mathbf{q},a)+   \mathbf{\Psi}^{(3)}(\mathbf{q},a)+\cdots.
\end{equation}

Here, $\mathbf{\Psi}^{(1)}$ corresponds to the Zeldovich approximation \citep{Zeldovich_1970} and captures the leading-order displacements. The second-order term, $\mathbf{\Psi}^{(2)}$, introduces non-linear, curl-free corrections, while $\mathbf{\Psi}^{(3)}$ includes higher-order contributions, capturing vorticity modes that are absent at lower orders \citep[e.g.][]{Bouchet_1994, Rampf_2012}. The peculiar velocity of the particles is given by $\mathbf{v}=a\,\dot{\mathbf{\Psi}}$. For a detailed derivation of these displacement fields, we refer to \citep{Zeldovich_1970,Bouchet_1992, Bouchet_1994,Rampf_2012}. 

 The codes used to generate initial conditions, such as those presented in \autoref{sec:ic_codes}, compute these displacement fields from the overdensity field. They then output particle positions and velocities in a format adequate to run a $N$‑body code.

\subsection{Primordial non‐Gaussianity}

While single-field slow-roll inflation predicts a nearly Gaussian primordial spectrum of perturbations, with departures of the order $\mathcal{O}(n_s - 1)$ \citep{Maldacena_2003, Creminelli_2004}, non-standard inflationary scenarios can produce observable non-Gaussianities \citep{Chen_2010, Byrnes_2010}.

Primordial non-Gaussianities (PNGs) are most commonly studied through their imprint on the three‐point function (or bispectrum) of the IC or primordial field. In this work, we focus on the local type of PNG, in which the non‐Gaussian potential is usually parametrised as \citep{Salopek_1990,Komatsu_2001}: 
\begin{equation}
\begin{aligned}
\phi_{\rm NG}(\mathbf{x})
= \phi_G(\mathbf{x}) &+ f_{\rm NL}\bigl(\phi_G^2(\mathbf{x}) - \langle\phi_G^2\rangle\bigr)\\
&+ g_{\rm NL}\bigl(\phi_G^3(\mathbf{x}) - 3\langle\phi_G^2\rangle\,\phi_G(\mathbf{x})\bigr),
\end{aligned}
\label{eq:phi_local_png}
\end{equation}
where $\phi_G$ is the Gaussian potential while the amplitudes $f_{\rm NL}$ and $g_{\rm NL}$ control the quadratic and cubic deviations respectively. 

Current constraints from the Planck temperature and polarisation data are $f_{\rm NL}^{\rm local} = -0.9 \pm 5.1\, (68\%\, {\rm c.l.}),\quad g_{\rm NL}^{\rm local} = (-5.8 \pm 6.5)\times10^4 \, (68\%\, {\rm c.l.})$ \citep{Planck_2018b}, close to the cosmic variance limit. Future improvements must come from large‐scale structure surveys, which can potentially reach sensitivities of $|f_{\rm NL}|\sim1$ \citep{Ballardini_2019,Heinrich_2024}, enough to rule out single‐field slow‐roll models (which predict $|f_{\rm NL}|\ll1$). 

A powerful probe of local PNG in the galaxy or halo clustering is the emergence of a scale‐dependent bias on large scales. In the presence of local PNG, the linear bias expansion of the halo overdensity field, $\delta_h$, acquires an extra term proportional to the primordial potential:
\begin{equation}
\delta_h
= b_1\,\delta_m + b_\phi\, f_{\mathrm{NL}}\phi\,.
\end{equation}
Here $b_1$ is the linear bias and $b_\phi$ is the bias parameter associated with the $f_{\rm NL} \phi$ operator. Physically, $b_\phi$ represents how halos respond to the presence of local PNG.  Using the Poisson equation (\autoref{eq:poisson_fourier}), one substitutes $\phi$ by $\delta_m$, which yields: 
\begin{equation}
P_{hh}(k,z)
= \bigl(b_1(z) + b_\phi(z)\,f_{\rm NL}\,\alpha(k,z)\bigr)^2 P_{mm}(k,z)\,,
\label{eq:pk_scale_dependent_bias}
\end{equation}
which exhibits an enhancement proportional to $k^{-4}$ on the largest scales. Consequently, larger survey volumes dramatically improve the sensitivity to measure $f_{\rm NL}$.

A key challenge is the degeneracy between $f_{\rm NL}$ and the bias parameter $b_\phi$, quantifying how the tracer responds to large‐scale potential fluctuations. Under the assumption of a universal halo mass function, that is, depending only on mass, the universality relation is predicted \citep{Dalal_2008,Matarrese_2008}:
\begin{equation}
b_\phi = 2\,\delta_c\,(b_1 - 1)\,.
\end{equation}
Here $\delta_c \simeq 1.686$ the spherical collapse threshold. However, numerical and analytic studies reveal deviations from this relation for specific halo samples \citep[e.g.][]{Slosar_2008,Grossi_2009,Hamaus_2011,Lazeyras_2023,Adame_2024} and for galaxies \citep[e.g.][]{Barreira_2020,Voivodic_2021,Marinucci_2023}. To capture the deviations from the universality relation, the following parametrisation is commonly used,
\begin{equation}
b_\phi = 2\,\delta_c\,(b_1 - p)\,,
\label{eq:bphi_p}
\end{equation}
where the PNG‐response parameter $p$ is measured from simulations and can depend on the tracer sample.

\subsection{Initial conditions for local‐PNG simulations}\label{sec:ICaliasing}

To generate initial conditions with local PNG, \autoref{eq:phi_local_png} is applied to the Gaussian potential before computing the LPT displacements. Alternative methods have been discussed in the literature. Convolving $\phi$ with a given kernel allows the generation of non-Gaussianities with an arbitrary bispectrum \citep{Wagner_2012,Scoccimarro_2012}. The local PNG case can be recovered for a specific choice of kernel.  

We note that local non-Gaussianities not only couple the long-wavelength perturbations with the short-wavelength ones but also change the depth of the gravitational potential wells. This can alter the collapse of structures, potentially varying the regime of validity of the LPT approximation and hence the conclusions from \citet{Michaux_2020} about pushing the initial redshift to lower values. Although LPT ceases to be formally valid after shell-crossing, their results indicated that one can still push to lower $z_{\rm start}$ if most of the volume remains in the perturbative regime. For example, in a simulation with a box size $L_{\rm box} = 1\,h^{-1}{\rm Gpc}$ and $N_{\rm part} = 1024^3$, the best agreement with the continuous limit occurs at $z_{\rm start} = 11.5$ with $3$LPT, for the configurations they tested.  Although the role of the initial redshift and resolution was already explored in \citet{Stahl_2024} for scale-dependent $f_{\rm NL}$ models, here we extend their analysis to pure local PNG models and we study higher-order and halo statistics, as well as how the LPT order may impact convergence.

Numerical aliasing arises because we represent continuous fields on a finite, discrete grid and must be accounted for to capture the modified potentials on a mesh accurately. The maximum frequency that can be represented on the grid, is the Nyquist wave number, defined as $k_{\rm Nyq} \equiv \frac{N}{2} k_{\rm f}$, where $N$ is the grid size, and $k_{\rm f}$ is the fundamental mode, $k_{\rm f} = \frac{2 \pi}{L_{box}}$. Real‐space operations (\autoref{eq:phi_local_png}) like $\phi_G^2$ (and $\phi_G^3$) correspond to convolutions in Fourier space which inject power up to $2\,k_{\rm Nyq}$ (or $3\,k_{\rm Nyq}$ for the cubic term). Any mode beyond $k_\mathrm{Nyq}$ cannot be represented on the grid and is assigned (aliased) into lower-k modes, that is, those with $k\leq k_{\mathrm{Nyq}}$ that the grid can represent. Because $\phi^2_G$  only reaches $2k_{\rm Nyq}$, while $\phi^3_G$ extends to $3\, k_\mathrm{Nyq}$, the specific intervals of contaminated (aliased) modes differ, and hence the mitigation strategy.

\section{Simulations} \label{sec:simulations}

We ran a suite of $N$-body simulations to study the generation of initial conditions for local PNG simulations and strategies to remove the PNG aliasing signal ({\it dealiasing}). The simulation suite consists of a set of $140$ low-resolution simulations, $N_{\rm part}=512^{3}$ and $L_{\rm box}=1\,h^{-1}\mathrm{Gpc}$, with varying initial conditions. We used the LPT with orders from $1$ to $3$ and initial redshifts $z_{\rm start}=99,\,49,\,24,\,11.5$. The LPT–$z_{\rm start}$ combinations we considered are listed in \autoref{tab:simulations}. We excluded combinations that previous works showed that do not improve accuracy. For example, previous studies showed that $1$LPT fails well before $z_{\rm start}=49$, and that $2$LPT can match the $3$LPT performance at high redshift, rendering $3$LPT unnecessary at $z_{\rm start}=99$.  All the simulations presented in this paper assume the Planck cosmology summarised in \autoref{tab:cosmology} \citep{Planck_2018}.

\begin{table}
\centering
\caption{Summary of the simulations ran with $N_{\rm part}=512^3$ and $L_{\rm box}= 1\; h^{-1}{\rm Gpc}$, used to study the effect of LPT order, initial redshift, and aliasing in the presence of local PNG.}

\begin{tabular}{c|l|l}
\hline
\multicolumn{1}{l|}{LPT order} & Initial redshift     & PNG-model \\ \hline
$1$ LPT                        & $z_{\rm start} = 99$   & \multirow{7}{*}{\begin{tabular}[c]{@{}l@{}}$f_{\rm NL},g_{\rm NL} = 0$\\ $f_{\rm NL} = 100$\\ $f_{\rm NL} = -100$\\ $g_{\rm NL} = 10^7$\\ $g_{\rm NL} = -10^7$\end{tabular}} \\ \cline{1-2}
\multirow{3}{*}{$2$LPT}        & $z_{\rm start} = 99$   &                                                                                                                                                                              \\
                               & $z_{\rm start} = 49$   &                                                                                                                                                                              \\
                               & $z_{\rm start} = 24$   &                                                                                                                                                                              \\ \cline{1-2}
\multirow{3}{*}{$3$LPT}        & $z_{\rm start} = 49$   &                                                                                                                                                                              \\
                               & $z_{\rm start} = 24$   &                                                                                                                                                                              \\
                               & $z_{\rm start} = 11.5$ &                                                                                                                                                                              \\ \hline
\end{tabular}

\label{tab:simulations}
\end{table}

\begin{table}
\centering
\caption{Cosmological and input parameters of all simulations.}
\begin{tabular}{lr}
\hline
$\Omega_{\rm m}$                              &    $0.3099$ \\
$\Omega_{\rm b}$                              &    $0.0489$ \\
$\Omega_\Lambda$                              &    $0.6901$ \\
$h \equiv H_0$/(100 km\,s$^{-1}$\,Mpc$^{-1})$ &    $0.6774$\\
$\sigma_8$                                    &    $0.8090$ \\
$n_{\rm s}$                                   &    $0.9682$ \\
\hline
$L_{\rm box}$                                 &    $1\, h^{-1}\,{\rm Gpc}$\\
$N_{\rm part}$                                &    $512^3, 1024^3$ \\
\hline
\end{tabular}
\tablefoot{The cosmology is fixed to the Planck 2018 best-fit values \citep{Planck_2018} (table 2.20 from  \citet{Planck2018Table}). Here, $L_\mathrm{box}$ denotes the comoving box size of each simulation, and $N_\mathrm{part}$ is the total number of particles in each simulation (see \autoref{sec:simulations}).}
\label{tab:cosmology}
\end{table}

Regarding the local PNG models, we considered $f_{\rm NL}=0,\pm 100$ and $g_{\rm NL}=0,\pm10^{7}$. As the simulation boxes are relatively small, lower $|f_{\rm NL}|$ values would weaken the PNG signal, making it harder to study. The local non‑Gaussianity of type $g_{\rm NL}$ is expected to produce a stronger aliasing effect than $f_{\rm NL}$ cosmologies. Given that $\phi\sim10^{-5}$ in the initial snapshot and that $f_{\rm NL} \phi ^2 \sim g_{\rm NL} \phi^3 \sim 10^{-8}$, a value of $g_{\rm NL}=\pm10^{7}$ produces a fractional correction in $\phi_{NG}$ of the same order as $f_{\rm NL}=\pm100$.

For every combination of initial redshift, $z_{\rm start}$, and LPT order, we ran simulations at each amplitude of local PNG ($f_{\rm NL}$ and $g_{\rm NL}$). Each configuration was executed in aliased and  de-aliased modes (see \autoref{sec:aliasing} for details). Every run also has two realisations with inverted phases, following \citet{Pontzen_2016}. To isolate systematic differences, we used \textsc{MonofonIC} \citep{Hahn_2020,Michaux_2020} to set all the initial conditions, and we fixed the same seed of white noise for all runs.

All the simulations presented here were evolved with the $N$‑body code \textsc{PKDGRAV3} \citep{Potter_2017, Alonso_2023}. This code uses the fast multipole method to compute the gravitational force. The algorithm implemented in the code provides a scaling $\mathcal{O}(N)$ and takes advantage of GPUs for additional speed while maintaining high accuracy \citep[see][for a comparison]{Garrison_2019}. We saved snapshots only at $z=z_{\rm start}$ and $z=0$, at which we performed our study. Halos within the simulations were identified with \textsc{Rockstar} \citep{Behroozi_2012}. This method is based on a friends-of-friends algorithm in the 6D phase space, using the positions and velocities of the particles. Only halos containing at least 20 particles were considered for our analysis.

\subsection{Removing the PNG aliasing signal}\label{sec:aliasing}

To assess the impact of the PNG aliasing signal, introduced in \autoref{sec:ICaliasing}, we ran two pairs of simulations for each combination of $z_{\rm start}$, LPT order, and $f_{\rm NL}$/$g_{\rm NL}$ value (see \autoref{tab:simulations}). We refer to each pair as aliased and dealised. For the aliased simulations, we computed directly the $\phi^2$ term in the grid (i.e. naïve convolution), while for the de-aliased simulations, we used the Orszag 3/2 rule. 

The Orszag3/2 rule \citep{Orszag_1971} has been previously used in the literature but not, to our knowledge, in the context of simulations with PNG. This method consists of: (i) padding the original Fourier grid with $N^3$ cells with zeros up to $(3/2\, N)^3$; (ii) computing the non-linear product $\phi^2$ at this higher-resolution; (iii) truncating the grid back to the original size of $N^3$.  The zero padding by a factor of $3/2$ in each dimension increases the memory footprint by $(3/2)^3\approx 3.4$, which is often prohibitive if the original grid is already large. More generally, computing $\phi_G^n$ without aliasing requires padding to $((n-1)/2)N$ in each dimension \citep{Patterson_1971}. 

This Orszag3/2 rule was already implemented within \textsc{MonofonIC} for computing the fields required for $2$LPT and $3$LPT \citep{Michaux_2020}. In \citet{Stahl_2023}, this technique was extended also for removing the PNG aliasing. Here, we introduce the possibility of disabling the Orszag3/2 convolver for the PNG part (while preserving it for the LPT part) for generating our dealised simulations. 

For the cubic term, $g_{\rm NL}$, we avoid the direct $2N$ zero padding by performing two sequential convolutions. Using the $3/2\; N$ padding, we first calculate $\phi_G^2$ and then $\phi_G^3$. Therefore, the maximum memory increase per pass is a factor $\left(3/2\right)^3\sim3.4$. This is well below the $2^3=8$ of a one‐shot cubic convolution, yet still free of aliasing.

\subsection{Initial condition codes} \label{sec:ic_codes}

To validate the robustness and accuracy of the primordial non-Gaussian implementation, we performed a controlled comparison of four state-of-the-art initial conditions (IC) generators: \textsc{MonofonIC}\footnote{\url{https://github.com/cosmo-sims/monofonIC}}\citep{Hahn_2020,Michaux_2020}, \textsc{2LPTPNG}\footnote{\url{https://github.com/dsjamieson/2LPTPNG/tree/main}}\citep{Scoccimarro_2012}, \textsc{FastPM}\footnote{\url{https://github.com/fastpm/fastpm}}\citep{Feng_2016}, and \textsc{LPICOLA}\footnote{\url{https://github.com/CullanHowlett/l-picola}}\citep{Howlett_2015}.  The last two codes are not dedicated IC generators but can act as such when instructed to write the initial snapshot. The LPT orders and PNG models available for each of the codes are summarised in \autoref{tab:codes}.

\begin{table}
\centering
\caption{Overview of the initial‐condition generation codes used in this work.}
\begin{tabular}{llll}
Code  & LPT & PNG   & References   \\ 
  & orders & models   &    \\ \hline
\textsc{MonofonIC} & 1,2,3       & $f_{\rm NL}^{\rm loc}$, $g_{\rm NL}^{\rm loc}$             & \citeauthor{Hahn_2020} \\
 &        &        & \citeauthor{Michaux_2020} \\
\textsc{2LPTPNG}   & 1,2         & $f_{\rm NL}^{\rm loc}$, $f_{\rm NL}^{eq}$, $f_{\rm NL}^{ortho}$ & \citeauthor{Scoccimarro_2012}        \\
\textsc{FastPM}    & 1,2         & $f_{\rm NL}^{\rm loc}$                                        & \citeauthor{Feng_2016}               \\
\textsc{LPICOLA}   & 1,2         & generic $f_{\rm NL}$                                          & \citeauthor{Howlett_2015}           
\end{tabular}
\tablefoot{We also include the supported LPT orders, implemented primordial non‐Gaussianity models, and key references. }
\label{tab:codes}
\end{table}

\textsc{MonofonIC} implements local PNG through a convolution in real space, which is corrected for aliasing with the Orszag3/2 technique (\autoref{sec:aliasing}). In contrast, the public version of \textsc{2LPTPNG} does not remove aliasing and resembles the naïve convolution in \textsc{MonofonIC} (\autoref{sec:aliasing}). \textsc{FastPM} also uses real-space convolution but applies a low-pass filter to remove small-scale contributions. The parameter \textsc{kmax\_primordial\_over\_knyquist} ($k_{\rm max}$ hereafter) controls at which scale this filter is applied. Finally, \textsc{LPICOLA} implements a different approach based on \citet{Scoccimarro_2012} that allows the generation of general PNG (if it is compiled with the \textsc{GENERIC\_FNL} flag). It can reproduce local PNG by choosing an appropriate kernel. The code also has the option to operate in the same way as \textsc{2LPTPNG} (compiling with the \textsc{LOCAL\_FNL} flag), but this is redundant for our comparison here.

We compared the four IC generators using identical settings to isolate how each code models the PNG signal. We generated 2LPT IC at $z_{\rm start}=49$ with $f_{\rm NL}=0,\pm 100$. These are the common settings available to all codes, as shown in \autoref{tab:codes}. We performed a total of $36$ simulations for this study, with $N_{\rm part}=1024^{3}$ particles and the same box size as for all the simulations presented in this work, $L_{\rm box} = 1\, h^{-1}\, \mathrm{Gpc}$.  We used a grid with the same $N=1024^{3}$ to compute the LPT fields for the ICs and fix this choice consistently for all codes. We also tested increasing this number to $2048^{3}$ and found that our results remain unchanged.

For \textsc{FastPM} we varied the value of the parameter $k_{\rm max}$, setting $k_{\rm max} = \frac{1}{4},\, \frac{2}{3}$ and $1$. \textsc{FastPM} applies a low-pass filter for the $\phi^2$ field at $k=k_{\rm max}\cdot k_{\rm Nyq}$. This filter removes the PNG signal beyond the frequency $k_{\rm max}\cdot k_{\rm Nyq}$, as shown in \autoref{fig:relerr_combined_codes}. This is discussed in \autoref{sec:results_ic_codes}.

After running the simulations, we found that the white noise \textsc{LPICOLA} produces in the Gaussian mode (\textsc{GAUSSIAN}) is different from the one it generates when running the code in the general PNG mode (\textsc{GENERIC\_FNL}) or the local PNG mode (\textsc{LOCAL\_FNL}). The other three codes considered here generate a white noise compatible with the Gaussian \textsc{LPICOLA}. This results in larger relative differences in all statistics measured for \textsc{LPICOLA} compared to the other codes. Nevertheless, as we show below, all the differences are compatible with the cosmic-variance limit.

\subsection{Reference simulations}\label{sec:reference_sims}

To study the impact of the parameter variations introduced above, we require a baseline set of runs that we refer to as the reference simulations. The reference simulations were run in a box side with side $L_{\rm box}=1\,h^{-1}\mathrm{Gpc}$ and $N_{\rm part}=1024^{3}$, providing a mass resolution eight times higher than for those runs with $N_{\rm part}=512^{3}$. We ran Gaussian reference simulations, as well as with PNG, for  $f_{\rm NL}=\pm100$ and $g_{\rm NL}=\pm10^7$. The ICs for all reference simulations were generated with \textsc{MonofonIC} and their main characteristics are summarised in \autoref{tab:reference_simulations}. The PNG aliasing signal was removed with the Orszag3/2 rule (see \autoref{sec:aliasing}) for all the reference simulations.

\begin{table}[]
\caption{Parameters used to set up the initial conditions for the reference simulations for the three studies presented in this work.}
\begin{tabular}{llll}
Reference simulations  &LPT & Initial & De-aliased  \\ 
in section & order & redshift  &   \\ \hline
\ref{sec:results_initial_z} & $3$    & $24$   & Yes      \\
\ref{sec:results_aliasing}  & $3$    & $24$   & Yes       \\
\ref{sec:results_ic_codes}  & $2$    & $49$   & Yes    
\end{tabular}
\tablefoot{All the reference simulations have been initialised using \textsc{MonofonIC}, with a box size of $L_{\rm box}=1\,h^{-1}\mathrm{Gpc}$ and a number of particles of $N_{\rm part}=1024^{3}$. We produce a pair of simulations for each PNG cosmology (i.e. $f_{\rm NL}=\pm100$ and $g_{\rm NL}=\pm10^7$). }
\label{tab:reference_simulations}
\end{table}

For the analysis of the initial redshift, LPT order, and aliasing in the presence of local PNG (\autoref{sec:results_initial_z} and \autoref{sec:results_aliasing}), we generated the initial conditions at $z_{\rm start}=24$ using $3$LPT. This choice, motivated by \citet{Michaux_2020}, balances accuracy with the risk of starting too late. 

For the comparison of IC generators (\autoref{sec:results_ic_codes}), our reference simulations have IC generated with \textsc{MonofonIC} assuming $2$LPT and $z_{\rm start}=49$. As the other codes cannot produce $3$LPT initial conditions, our IC set-up is common to all the codes under study. This allows us to analyse the difference in the PNG implementation between the codes. We chose \textsc{MonofonIC} as our reference code because we have tested its convergence in \autoref{sec:results_initial_z} and \ref{sec:results_aliasing}. Moreover, \textsc{MonofonIC} is the only code that supports dealiasing via the $3/2$ zero-padding (see \autoref{sec:aliasing}) among those considered here.

\subsection{Measuring the power spectrum and bispectrum}\label{sec:pkb}

We computed the power spectra with the \textsc{Pylians} package\footnote{\url{https://pylians3.readthedocs.io/en/master/}} \citep{Pylians}. We assigned dark‑matter particles onto a $1024^{3}$ Cartesian density grid using a piecewise cubic spline (PCS) mass assignment scheme. Then, we measured the Fourier‑space density contrast on that grid. With PCS, the aliasing correction is well understood, and the resulting estimate remains accurate up to the Nyquist wave number, $k_{\rm Nyq}$:
\begin{equation}
    k_{\rm Nyq} = \frac{\pi N_{\rm part}^{1/3}}{L_{\rm box}}
\end{equation}

The same density grid was used for the bispectrum analysis. We used the GPU‑accelerated code \textsc{BFast}\footnote{\url{https://github.com/tsfloss/BFast/tree/main}} to evaluate all triangle configurations in bins of width $\Delta k = 3k_{f}$ up to $k_{\rm max} = \frac{2}{3} k_{\rm Nyq}$ to avoid aliasing associated with the bispectrum estimator itself \citep{Sefusatti_2016}. In our analyses, we focused on the squeezed limit of the bispectrum $(k_1\sim k_2 \ggg k_3)$, as this is the configuration in which most of the local PNG signal resides.

\section{Results}\label{sec:results}

In this section, we present how the variation of initial redshift, LPT order, and local PNG aliasing influences the accuracy of $N$-body simulations. We focus on the convergence of the matter power spectra, bispectra, and halo statistics. Our goal is to identify which simulation set-up and IC generation approach yields the most accurate results in the presence of local PNG for a given computational cost.

\subsection{Initial redshift and LPT order} \label{sec:results_initial_z}

\citet{Michaux_2020} demonstrated that starting $N$-body simulations at a lower redshift (i.e. a later start) and using higher-order LPT schemes improve their convergence towards the continuous limit. 

We checked whether these findings also hold for local PNG. Positive $f_{\rm NL}$ or $g_{\rm NL}$ skews the one-point density distribution, enhancing the abundance of high-density regions, potentially causing earlier shell-crossing and narrowing LPT's valid range. Conversely, negative $f_{\rm NL}$ or $g_{\rm NL}$ suppresses the high-density tails, delaying collapse and allowing a lower $z_{\rm start}$. To investigate this, we performed simulations that varied the LPT order, initial redshift, and PNG parameters, as summarised in \autoref{tab:simulations}.

\begin{figure*}[!hbt]
  \centering
  \begin{subfigure}[b]{0.45\textwidth}
    \centering
    \includegraphics[width=\linewidth]{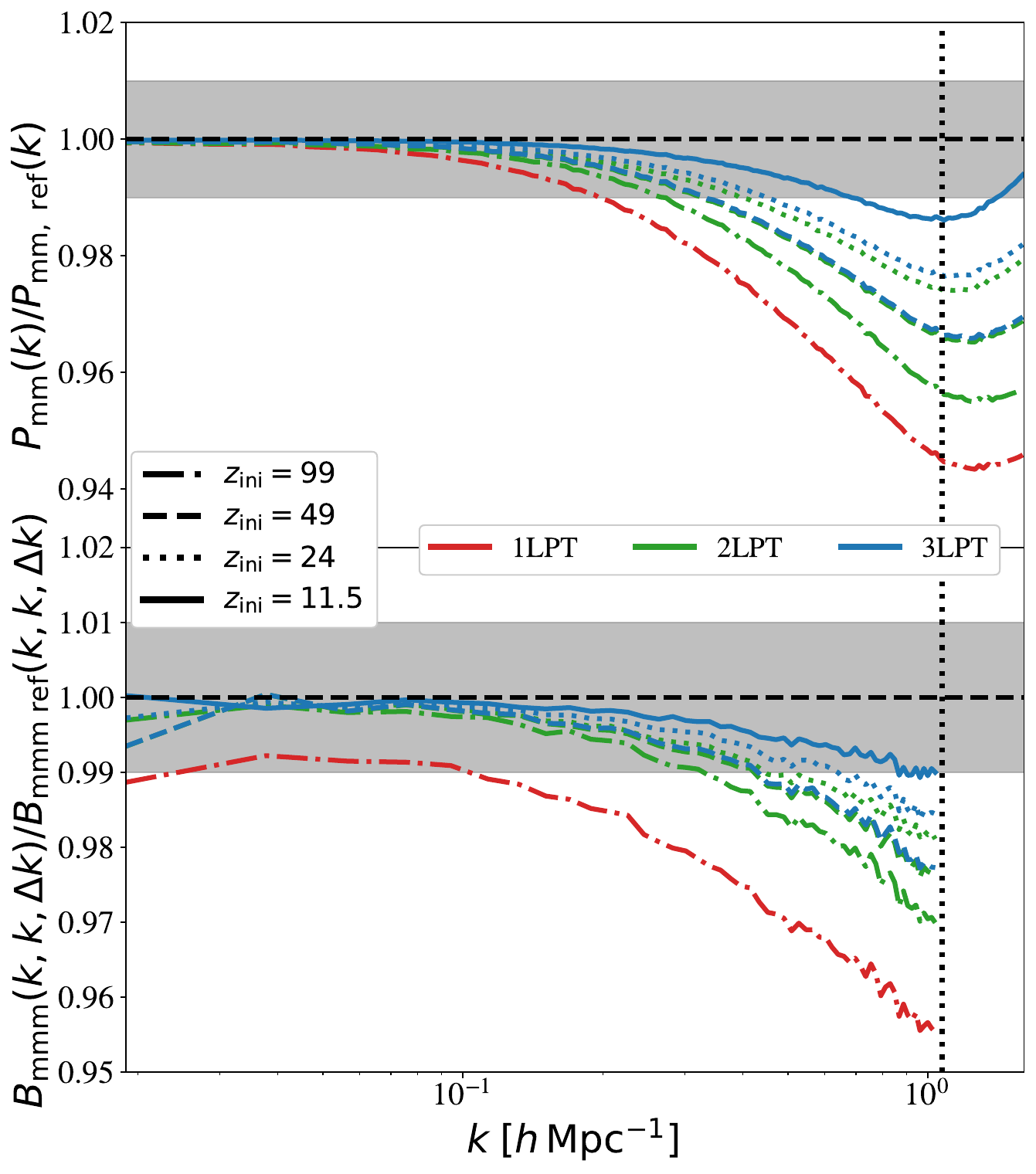}
    \caption{$f_{\rm NL} =0 $ }
    \label{fig:pk_bk_zini_lpt_fnl0}
  \end{subfigure}
  \hfill
  \begin{subfigure}[b]{0.45\textwidth}
    \centering
    \includegraphics[width=\linewidth]{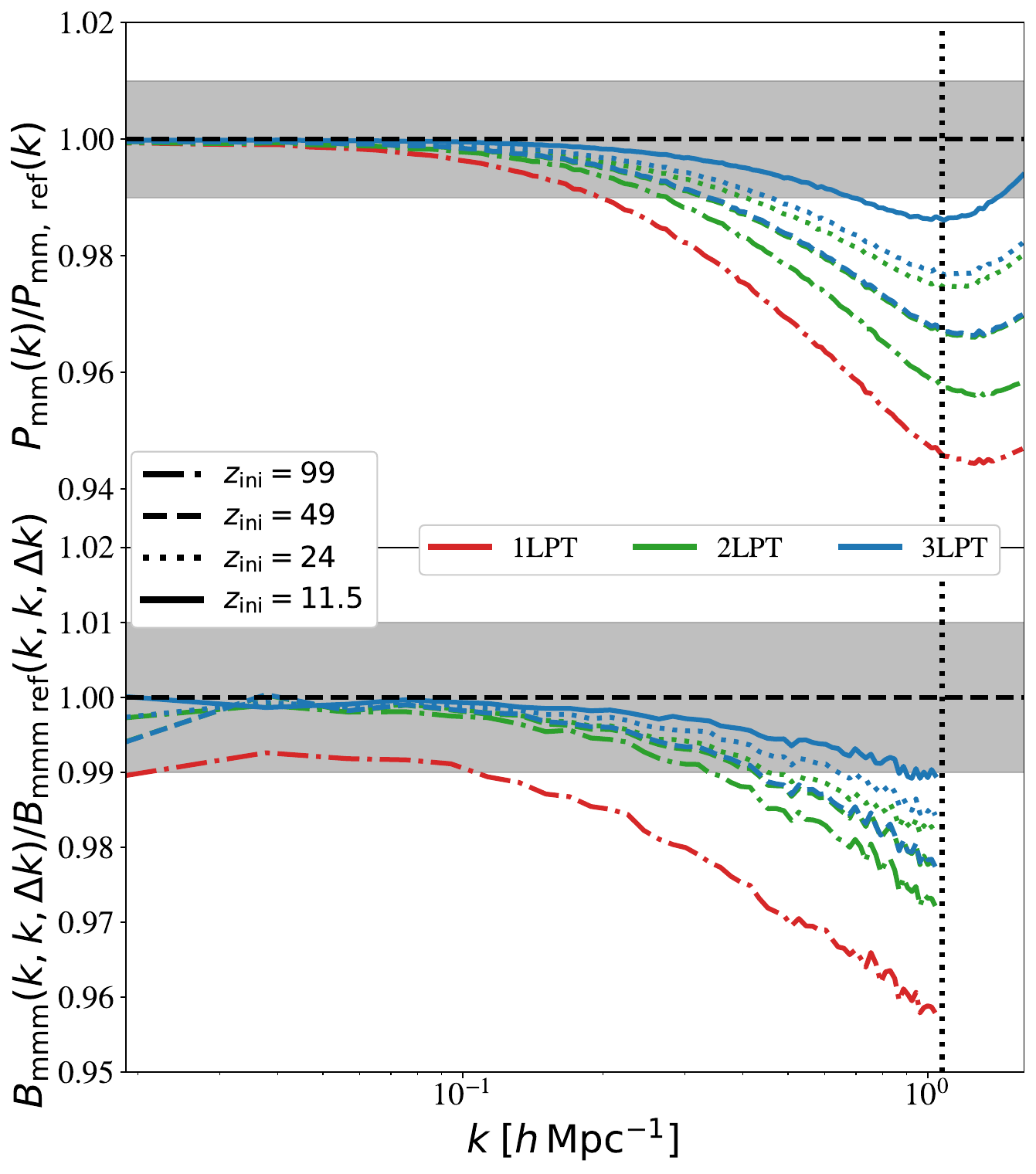}
    \caption{$f_{\rm NL}=100$}
    \label{fig:pk_bk_zini_lpt_fnl100}
  \end{subfigure}
  \caption{Comparison at $z=0$ of the matter power spectrum (upper panels) and bispectrum (lower panels, squeezed limit $(k_1 = k_2 = k;\; k_3=3k_f)$) with respect a high-resolution reference simulation (see \autoref{sec:reference_sims}).  The left panel shows the results for $f_{\rm NL}=0$ simulations, and the right panel for those with $f_{\rm NL} = 100$. The line colour encodes the LPT order used for the initial conditions (red: $1$LPT, green: $2$LPT, and blue: $3$LPT), while the line styles indicate the initial redshift. The shaded regions indicate the $\pm 1\%$ around the reference simulation. The results are nearly identical for $f_{\rm NL} = 0$ and $f_{\rm NL} = 100$. A later start using higher-order LPT schemes improves the agreement with the higher-resolution reference simulations.}
  \label{fig:pk_bk_zini_lpt}
\end{figure*}

Varying the initial redshift and the order of the LPT schemes can affect the power spectrum and the bispectrum, as shown in \autoref{fig:pk_bk_zini_lpt}, where for the bispectrum, we show the squeezed limit $(k_1 = k_2 = k,\; k_3 = 3k_f)$. The characteristics of the simulation used as a reference are discussed in \autoref{sec:reference_sims}.  All power spectra and bispectra are shown in relation to that reference unless otherwise noted.

\subsubsection{Power spectrum and bispectrum}

For the power spectrum in the Gaussian case ($f_{\rm NL}=0$), all starting redshift and LPT configurations agree at large scales ($k < 0.1\,h\,{\rm Mpc}^{-1}$) but show a power suppression at smaller scales. This attenuation is stronger for higher $z_{\rm start}$, up to a $5\%$ at $k=1\,h{\rm Mpc}^{-1}$. Higher-order LPT realisations (e.g. $2$LPT and $3$LPT) at $z_{\rm start}=49$ converge to the same solution but are biased low compared to the reference. Because both cases show the same small-scale deficit relative to the reference, the effect cannot be traced to truncated perturbation theory but instead points to a discreteness-driven origin \citep{Michaux_2020,Joyce_2005}. 

A similar pattern emerged for the squeezed-limit bispectrum (lower panels of \autoref{fig:pk_bk_zini_lpt}), although it was more sensitive to the LPT order. The runs with $1$ LPT remain systematically biased, while switching to higher-order LPT improves the agreement, as expected \citep{McCullagh_2016,Baldauf_2015}. We have focused on the squeezed limit as is the configuration that local PNG induces \citep[e.g.][]{Bartolo_2004}, although the equilateral configuration $(k_1 = k_2 = k_3 =  k )$ showed $1-2\%$ larger deviations. These findings were consistent with \citet{Michaux_2020} in the Gaussian case.

We then performed the same analysis for $f_{\rm NL} = \pm 100$ and $g_{\rm NL} = \pm 10^7$ to check whether the local PNG changes this picture. The right panel in \autoref{fig:pk_bk_zini_lpt} shows the $f_{\rm NL} = 100$ result. As in the Gaussian case, the convergence improved when starting the simulations at lower redshifts and employing higher-order LPT. Similar behaviour appeared for all other PNG parameters we tested, indicating that the results from \citet{Michaux_2020} still hold for moderate to high $f_{\rm NL}$ or $g_{\rm NL}$. 

We note that all our conclusions are based on runs with $|f_{\rm NL}|=100$ and $|g_{\rm NL}|= 10^7$. We expect that these conclusions also hold for smaller non-Gaussian amplitudes as they approach the Gaussian case, where this is also valid. However, larger $f_{\rm NL}/g_{\rm NL}$ values may require more testing.

\subsubsection{Halo mass function}

\begin{figure}
\centering
\includegraphics[width=\columnwidth]{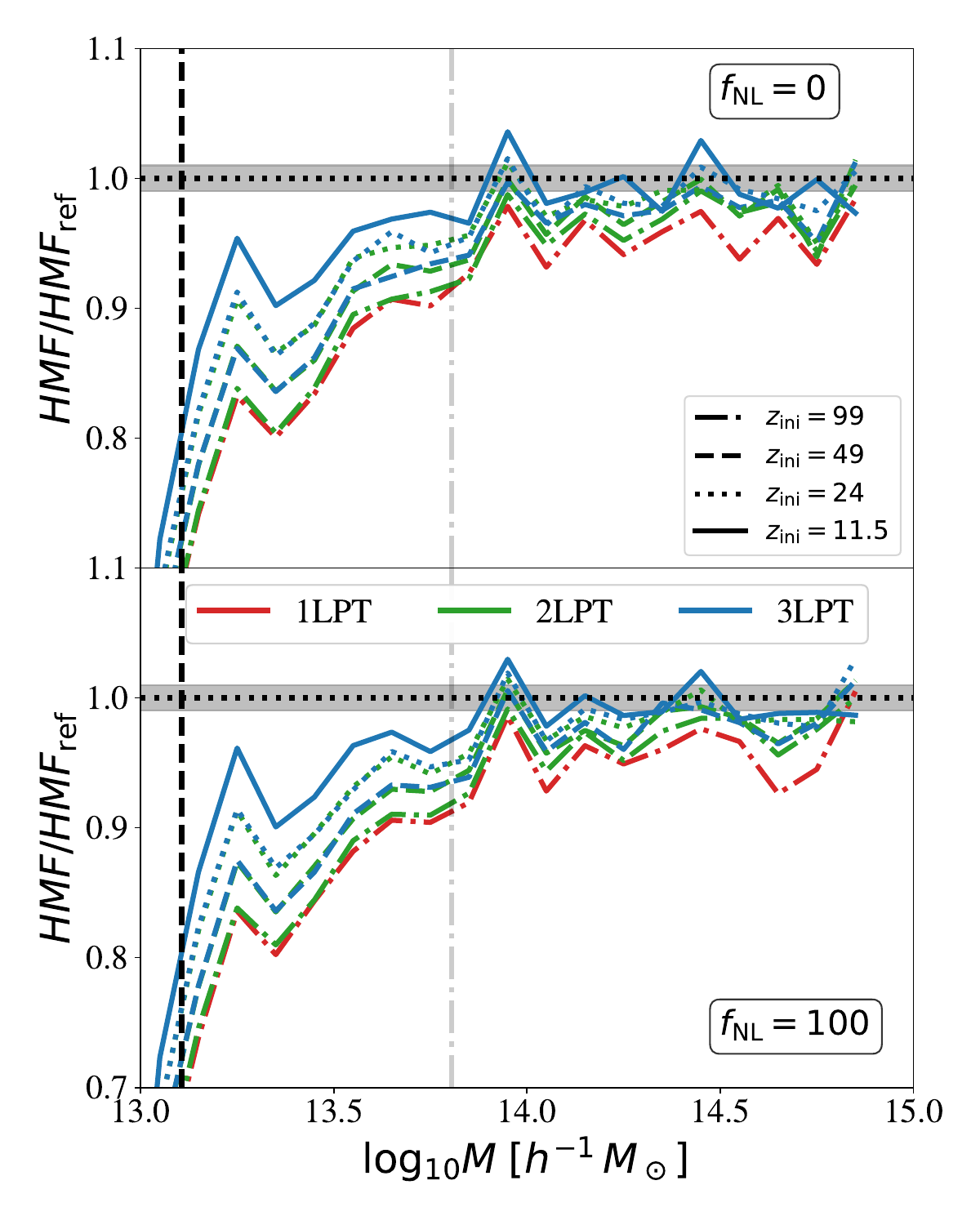}
\caption{Comparison at $z=0$ of the halo mass function for different configurations of the initial conditions (see \autoref{sec:simulations}) with respect to the higher-resolution reference simulations (see \autoref{sec:reference_sims}). The upper panel shows the results for the $f_{\rm NL}=0$ simulations, while the lower panel shows the case with $f_{\rm NL}=100$. The line colour encodes the LPT order used for the initial conditions (red: $1$LPT, green: $2$LPT, and blue: $3$LPT), while the line styles indicate the initial redshift. The dashed vertical lines show the limit of $20$ particles per halo; the limit of $100$ is indicated by the dash-dotted vertical lines. The shaded regions indicate the $1\%$ difference around the reference simulation. Starting the simulation at lower redshifts improves the convergence towards the high-resolution reference. }%
\label{fig:hmf_zini_lpt}
\end{figure}

Next, we examine the impact on halo populations. We computed the halo mass function (HMF) in the bins $\Delta \log M = 0.1$, shown in \autoref{fig:hmf_zini_lpt}. All runs showed a lack of low-mass halos at the level $10\%- 20\%$  for $M_{\rm part} < 3 \times 10^{13} h^{-1} M_{\odot}$. This is expected since the halos containing fewer than $\sim 100$ particles are not fully resolved \citep{Power_2003,Mansfield2021}. However, the deficit was less severe at later starts, regardless of the LPT order. In contrast, very early starts (e.g.$z_{\rm start} = 99$) introduced a systematic $\sim 4\%$ bias across the entire mass range. Likewise, $3$LPT at $z_{\rm start} = 11.5$ best matched the reference of high resolution. Our local PNG runs ($f_{\rm NL}= \pm 100$, $g_{\rm NL} = \pm 10^7$) revealed the same picture, confirming that these recommendations are robust to moderate values of $f_{\rm NL}$ or $g_{\rm NL}$.

These results reflect two competing effects. First, higher-order LPT schemes (e.g. 3LPT) provide more accurate displacements by including additional perturbative corrections, which formally improve convergence towards the fluid limit before shell-crossing. This is expected from theory, as the LPT expansion converges in powers of the linear growth factor $D(z)$. However, at very early times (e.g. $z_{\rm ini} = 49$), $D(z)$ is small, so even low-order LPT (such as 2LPT) closely approximates the full solution, explaining why 2LPT and 3LPT are nearly indistinguishable in that regime. Second, discreteness effects due to the initial particle lattice become more prominent when starting too early \citep{Michaux_2020}. These can be suppressed by initialising at lower redshifts, where the particle distribution has evolved away from the initial grid symmetry, and hence motivating the use of higher-order LPT schemes.

Lagrangian perturbation theory is only valid before shell-crossing. After this point, higher-order terms can actually diverge more rapidly from the true solution. Despite this limitation, our results show that even at very low starting redshifts (e.g. $z_{\rm start} = 11.5$), 3LPT yields better agreement with high-resolution simulations than any of the earlier-start configurations we tested. These results might change for simulations with a different mass resolution.

\subsection{Aliasing in local PNG simulations}\label{sec:results_aliasing}

Inducing local PNG in the initial conditions requires computing $\phi_G^2$, or $\phi_G^3$ in \autoref{eq:phi_local_png}. This introduces spurious high-frequency signals that alias onto the grid if not corrected (see \autoref{sec:ICaliasing}). This PNG aliasing increases power, in particular on small scales. However, to our knowledge,  this effect has not been measured before.

We performed two sets of simulations, aliased and de-aliased, to quantify the PNG aliasing signal. In the de-aliased reference simulations, the PNG aliasing signal was removed with the Orszag 3/2 rule (\autoref{sec:aliasing}).  In this subsection, we show only the results for simulations started with $3$LPT and $z_{\rm start}=24$ unless said otherwise.

\subsubsection{Power spectrum}

In the left panel of \autoref{fig:relerr_combined}, we present the relative differences in the matter power spectrum between the simulations with aliasing compared to the ones corrected for this effect (which in the figures appear labelled as dea). In the upper panel, we show the results of the initial redshift $z_{\rm start}=24$. We find that both $f_{\rm NL}$ and $g_{\rm NL}$ cosmologies generate a spurious signal when the aliasing effect is not corrected. The differences at large scales are negligible, $\leq 0.01\%$ for $g_{\rm NL}$ cosmologies and $\leq 10^{-5}\%$ for $f_{\rm NL}$ cosmologies with respect to the de-aliased reference. At smaller scales, the PNG aliasing signal reaches $\sim 3\%$ for $g_{\rm NL}$ cosmologies, while for $f_{\rm NL}$ it remained $\leq 0.1\%$ up to the Nyquist frequency. 

We also studied the power spectrum at $z=0$. (\autoref{fig:relerr_combined}). For the $g_{\rm NL}$ cosmologies, the relative differences of the aliased simulations with respect to the de-aliased one at small scales ($k\gtrsim 0.7\,h\,\mathrm{Mpc}^{-1} $) are smaller at final redshift than at initial redshift ($0.1\%$). In contrast, at large scales, they remain at a similar level ($0.01\%$) at either redshift.

We found a similar behaviour for $f_{\rm NL}$ cosmologies. The relative differences at the Nyquist frequency were suppressed to $0.01\%$ at late times. We found an increase of $0.001\%$ in the relative differences on large scales. We argue that this is not an effect of the aliasing itself but rather due to floating-point roundoff errors.

To explicitly check this, we also ran a set of simulations using the different convolvers for the $f_{\rm NL}=0$ case (black lines). Both realisations, aliased and de-aliased (see \autoref{sec:aliasing}), should return exactly the same initial conditions, as this effect only is sourced if $f_{\rm NL}$ or $g_{\rm NL}$ differ from zero. However, small numerical errors arise from the float point number operations, leading to slightly different initial conditions. This is shown in the upper left panel of \autoref{fig:relerr_combined}, where we observe some fluctuations in the power spectrum at $10^{-6}\%$ level. Then, after evolving to $z=0$ (lower left panel), these tiny differences amplify to the $10^{-3} \,\%$ level. Precisely, these fluctuations match what we found for the $f_{\rm NL} \neq 0$ cosmologies at $z=0$.  

\subsubsection{Bispectrum}

\begin{figure*}
  \centering
  \includegraphics[width=0.7 \linewidth]{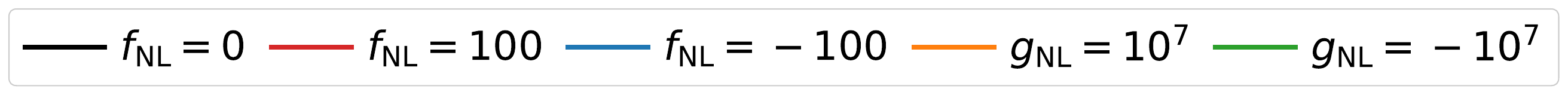}\\[-0.8ex]
  \begin{subfigure}[t]{.495\linewidth}
    \centering
    \includegraphics[width=\linewidth]{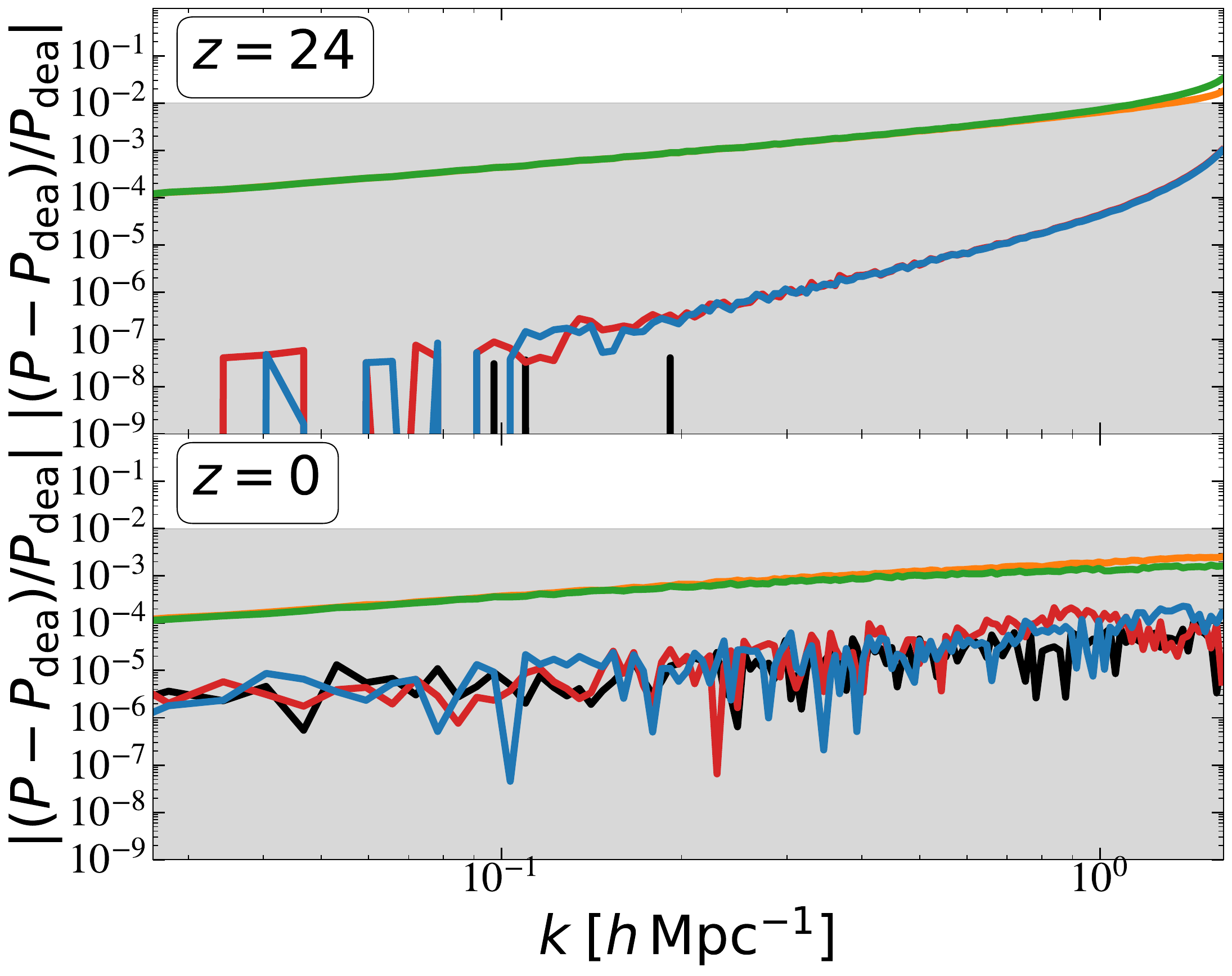}
    \caption{Power spectrum}\label{fig:pk_aliasing}
  \end{subfigure}
  \hfill
  \begin{subfigure}[t]{.495\linewidth}
    \centering
    \includegraphics[width=\linewidth]{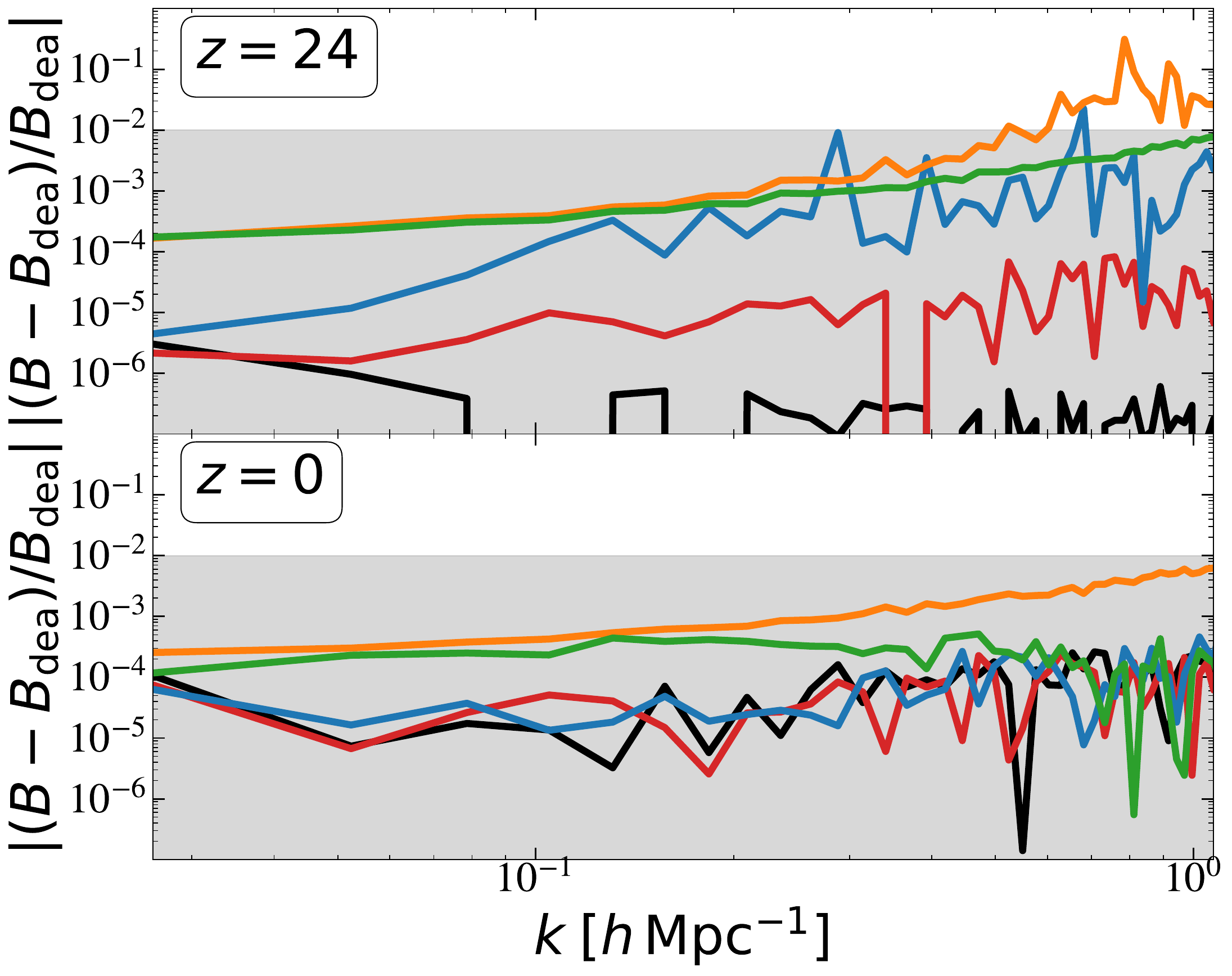}
    \caption{Bispectrum  $(k_1=k_2= k;\;k_3=k_f )$}\label{fig:bk_aliasing}
  \end{subfigure}
  \caption{Relative difference of the aliased with respect to the de-aliased (see \autoref{sec:aliasing}) matter power spectrum (left panel) and bispectrum in the squeezed limit $(k_1=k_2= k;\;k_3=k_f )$ (right panel). In the upper panels, we show the differences for initial conditions at $z=24$, and in the lower panels, the results for the final snapshot at $z=0$. The results for simulations with different PNG models are shown in different colours, as indicated in the legend. The shaded areas show the $1\%$ difference around the higher-resolution reference simulation (\autoref{sec:reference_sims}). Aliasing introduces a spurious signal in the initial conditions (upper panels) for both the power spectrum and bispectrum, reaching $3\%$ and $10 \%$ differences, respectively, in the $g_{\rm NL}=10^{7}$ case. By late times ($z=0$), these discrepancies are suppressed to below $1 \%$ across all scales and PNG parameter values considered.}
  \label{fig:relerr_combined}
\end{figure*}

We extended the previous analysis to the bispectrum in the right panel of \autoref{fig:relerr_combined}. We focus our analysis on the squeezed limit of the bispectrum $(k_1\sim k_2 \ggg k_3)$  Although not shown explicitly, we also analysed other configurations of the bispectrum, such as the equilateral limit $(k_1\sim k_2 \sim k_3)$, reaching the same conclusions as here. 

At $z=z_{\rm start}$, as in the case of the power spectrum, we found a spurious signal in the aliased simulations (right panel of \autoref{fig:relerr_combined}). This effect was more prominent compared to what we saw for the power spectrum in the left panel of \autoref{fig:relerr_combined}, reaching, for some configurations of the bispectrum, a relative difference of $10\%$ for the $g_{\rm NL} = 10^7$ case and $5\%$ for $f_{\rm NL}=-100$, with respect to our reference de-aliased simulations. However, for $k<0.3\,h{\rm Mpc}^{-1}$, relative differences were always below $1\%$. We also measured this relative difference for Gaussian cosmology $(f_{\rm NL}=0)$ where the aliased and de-aliased simulations should be identical. We found that at the level of initial conditions, there are already some fluctuations at the level of $\lesssim10^{-6}$ due to numerical errors. 

When we analysed the results for the bispectrum at $z=0$, we again found the same picture that we found for the power spectrum: the aliasing signal is suppressed at small scales, remaining a sub-percent effect for all the cosmologies we explored. Moreover, the results for $f_{\rm NL}$ cosmologies are compatible with the Gaussian one at all scales.

\subsubsection{Halo mass function}

\begin{figure}
\centering
\includegraphics[width=\columnwidth]{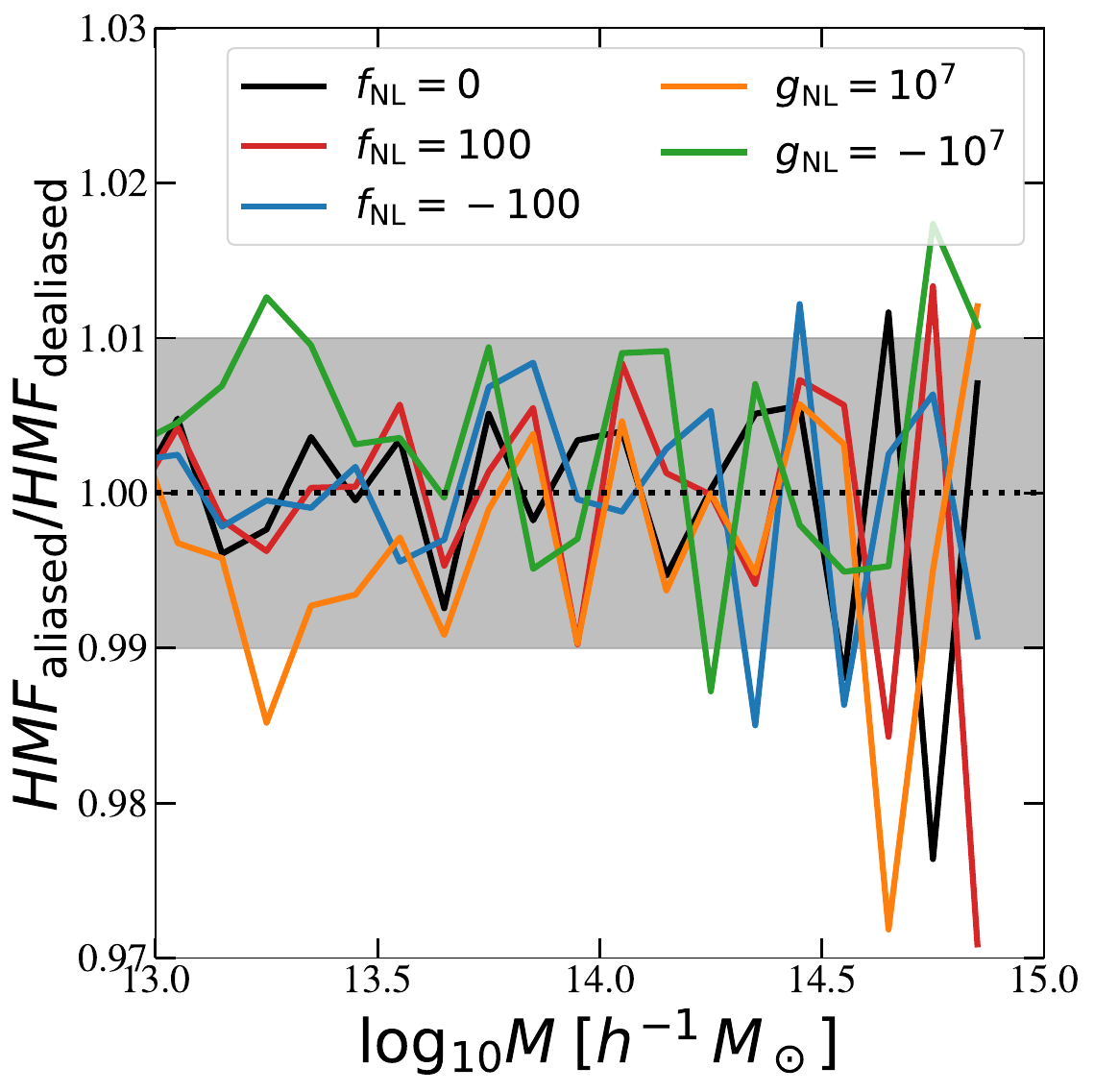}
\caption{Ratio at $z=0$ of the halo mass functions from simulations affected by PNG aliasing to the de-aliased corresponding simulation (see \autoref{sec:aliasing}). As indicated in the legend, the line colours correspond to the different PNG models used in the simulations. The shaded area represents the $1\%$ variation around the de-aliased simulations. The PNG aliasing signal produces variations below $3\%$ in the halo mass function at $z=0$.}%
\label{fig:hmf_aliasing}
\end{figure}

We now turn our attention to the halo mass functions, following the same procedure as in \autoref{sec:results_initial_z}. \autoref{fig:hmf_aliasing} shows the ratio of the halo mass functions for the different cosmologies, comparing the simulations where we have the aliasing effect with respect to the de-aliased simulations. The gravitational evolution and the process of halo finding are very non-linear, so even the tiny differences we saw in the initial conditions get amplified. This makes the halo mass function much noisier compared, for example, to the matter power spectrum that we have analysed previously. Nevertheless, we find an excellent agreement for all the cosmologies and mass ranges that are resolved with our simulations. 

All the mass functions are compatible within a $3\%$ level. This is true even for the largest mass bin, which we expect to be dominated by shot noise.

\subsubsection{Halo bias}

Finally, we studied the halo power spectrum and bispectrum, but this time focusing only on the $f_{\rm NL}$ cosmologies. These statistics are much noisier than the matter ones due to the non-linear nature of the gravitational collapse and halo finding, as in the HMF case. In order to extract more robust conclusions, we chose to compress these statistics into $2$ parameters, the halo linear bias $b_1$ and the PNG response parameter $p$ (defined in \autoref{eq:bphi_p}). To do this, we measured the halo power spectrum in the same way as we described in \autoref{sec:results} and fitted to \autoref{eq:pk_scale_dependent_bias} using a least squares approach. We followed the same procedure as in \cite{Avila_2022} to match the $f_{\rm NL}=0$ realisation with the $f_{\rm NL}=100$ (and similarly for $f_{\rm NL}=-100)$ in order to suppress the variance in the measurement of $b_\phi f_{\rm NL}$ and get more robust measurements of these parameters. We got the values for $b_\phi$ and then for $p$ by assuming the values of $f_{\rm NL}$ used to generate the initial conditions of the simulations.

\begin{figure}
\centering
\includegraphics[width=\columnwidth]{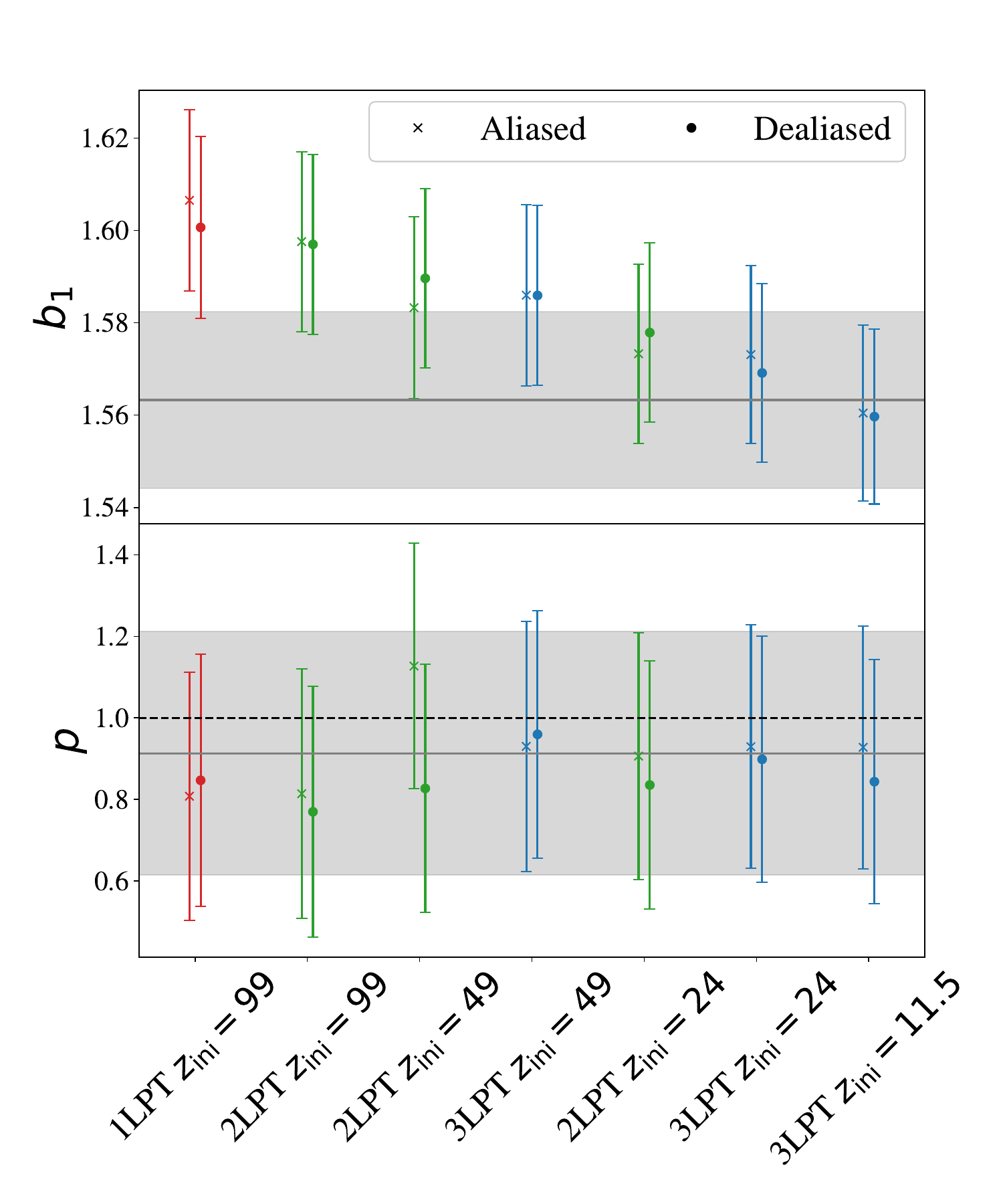}
\caption{Measurements at $z=0$ of the linear bias parameter $b_1$ (upper panel) and the PNG-response parameter $p$ (lower panel) for halos from the simulations summarised in \autoref{tab:simulations}. The colours indicate the LPT order used for the initial conditions (red: $1$LPT, green: $2$LPT, and blue: $3$LPT). The crosses present the best-fit values with the $1\sigma$ error bars from the fits for the simulations with a PNG aliasing signal. The circles present the results after removing the PNG alias signal, or de-aliased simulations (see \autoref{sec:aliasing}). The horizontal grey lines show the measurements from our reference high resolution simulations (see \autoref{sec:reference_sims}). The grey regions show the $1\sigma$ uncertainty from the fit to the reference simulation (\autoref{sec:reference_sims}). In the lower panel, the dashed horizontal line shows the universality relation,  $p=1$. }%
\label{fig:b1_p_aliasing}
\end{figure}

The halo linear bias $b_1$ and the PNG response parameter $p$ (defined in \autoref{eq:bphi_p}) from the simulations summarised in \autoref{tab:simulations}, both aliased and de-aliased (see \autoref{sec:aliasing}) are shown in \autoref{fig:b1_p_aliasing}. We find that aliasing had no significant impact in either of the parameters, $b_1$ or $p$. This aligns with the findings that aliasing has a small effect at late times.  The only configuration that shows a larger deviation for the $p$ parameter (although still compatible within the $1\sigma$ error bars) is $2$LPT with $z_{\rm start}=49$. A more detailed analysis of this configuration was carried out by varying the scale cuts and the mass range considered. When applying these changes, we found that everything aligned with our observations for the remaining configurations, leading us to conclude that this was a statistical fluctuation.

We find that earlier starts led to an artificially larger value of $b_1$, leading to a $2\sigma$ deviation for $z_{\rm start}=99$ when compared with our reference high-resolution simulation. These results support our findings in \autoref{sec:results_initial_z}, showing that later starts with higher-order LPT improved the convergence towards the higher-resolution simulations. Regarding the $p$ response parameter, we found that the LPT order and the initial redshift had a negligible impact.

\subsection{Comparison of state-of-the-art IC-codes} \label{sec:results_ic_codes}

We compared several public codes to generate local PNG initial conditions, namely \textsc{2LPTPNG}, \textsc{LPICOLA}, and \textsc{FastPM}. To isolate differences in their PNG implementations, we initialised all runs at $z_{\rm start}=49$ with $2$LPT and focused on $f_{\rm NL}$ as we discussed in \autoref{sec:ic_codes}. Although we only show results for $f_{\mathrm{NL}}=100$, we found similar results for $f_{\mathrm{NL}}=-100$.

\subsubsection{Power spectrum}
\begin{figure*} [!hbt]
  \centering
  \includegraphics[width=0.8 \linewidth]{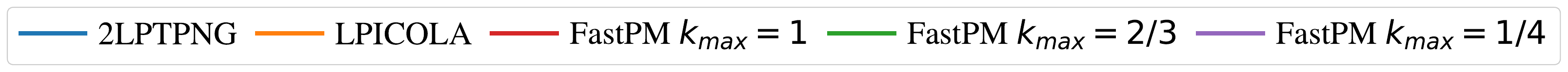}\\[-0.8ex]
  \begin{subfigure}[t]{.495\linewidth}
    \centering
    \includegraphics[width=\linewidth]{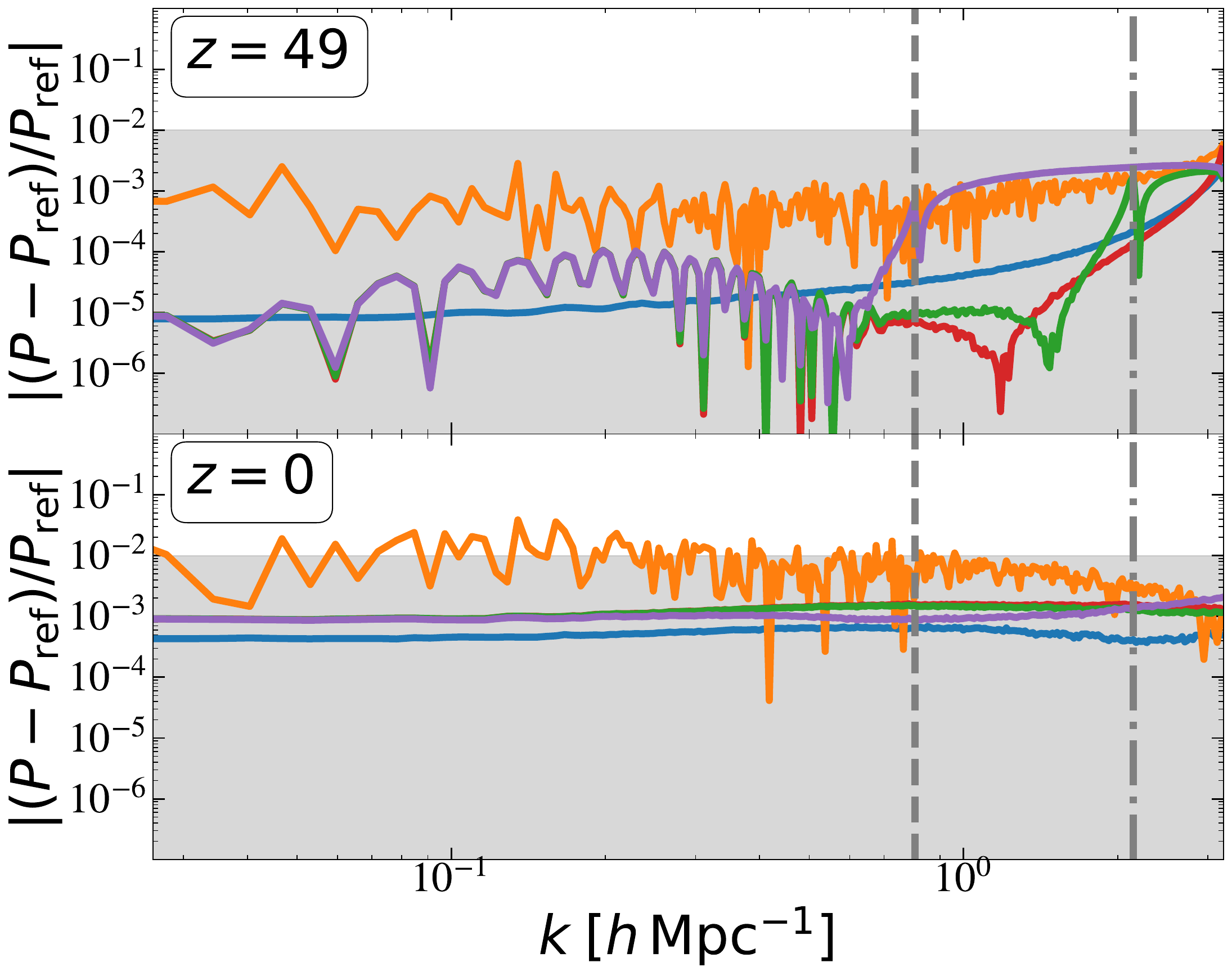}
    \caption{}\label{fig:pk_codes_ics}
  \end{subfigure}
  \hfill
  \begin{subfigure}[t]{.495\linewidth}
    \centering
    \includegraphics[width=\linewidth]{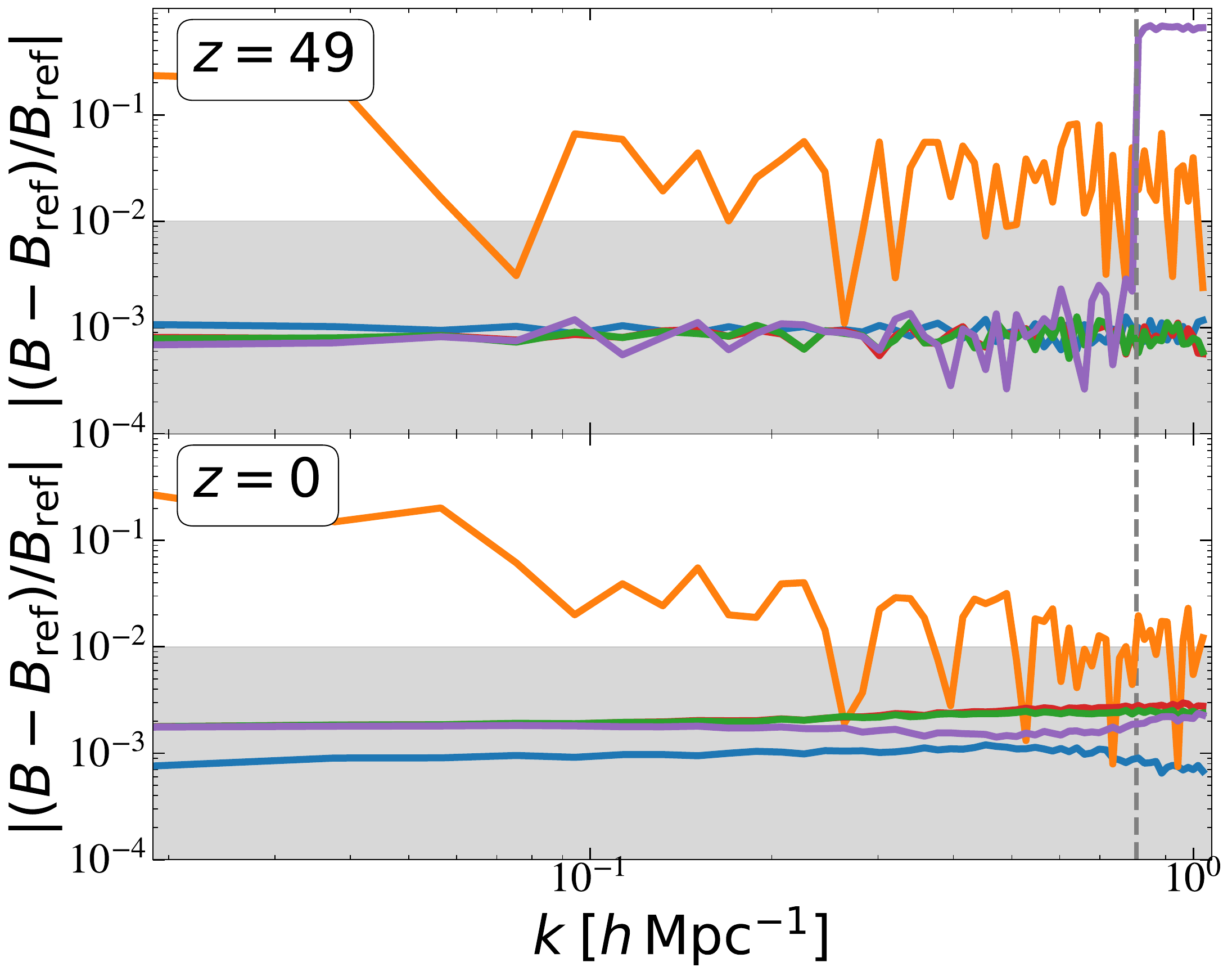}
    \caption{}\label{fig:bk_codes_ics}
  \end{subfigure}
  \caption{Relative variations of the matter power spectrum (left panel) and bispectrum in the squeezed limit $(k_1 = k_2 = k;\; k_3 = 3k_f)$ (right panel) with respect the reference simulation using \textsc{MonofonIC} (see \autoref{sec:reference_sims}). In the upper panels, we show the variations at the initial redshift, $z=49$, and in the lower panels at $z=0$. The codes are shown in different colours: \textsc{2LPTPNG} (blue), \textsc{LPICOLA} (orange), \textsc{FastPM} $k_{\rm max} = 1$ (red), \textsc{FastPM} $k_{\rm max} = 2/3$ (green), and \textsc{FastPM} $k_{\rm max} = 1/4$ (purple). The vertical dashed lines show $k = k_{nyq} \times 1/4$, and the vertical dash-dotted lines $k = k_{nyq} \times 2/3$. The grey shaded regions represent $1\%$ differences with respect to the reference. \textsc{LPICOLA} shows the largest variations at small $k$, due to a different realisation of the white noise than the reference simulation (see \autoref{sec:ic_codes}). Notably, \textsc{FastPM} with $k_{\rm max}=1/4$ shows a $80\%$ deviation for $k>\frac{1}{4} k_{nyq}$ for the bispectrum at the initial conditions (see \autoref{sec:results_ic_codes}). \textsc{2LPTPNG} and \textsc{FastPM}, with $k_{\rm max}=1$ and $2/3$, show an excellent agreement with respect \textsc{MonofonIC}, with variations below $\sim 0.1\%$ in the power spectrum and bispectrum at all scales, for both the initial conditions and at $z=0$.}
  \label{fig:relerr_combined_codes}
\end{figure*}

The left panel of \autoref{fig:relerr_combined_codes} shows the relative differences of the power spectrum with respect to \textsc{MonofonIC} at the initial redshift, $z=49$, and at $z=0$. \textsc{2LPTPNG} and \textsc{FastPM} with the configuration of $k_{\rm max}=1$ are the codes that yielded a better agreement with respect to our reference simulation, with relative differences of $0.001\%$ at large scales and less than $1\%$ even at the $k_{\rm Nyq}$. We noted that this increase in the differences at small scales is similar to the behaviour of the aliasing signal (left panel of \autoref{fig:relerr_combined}). Neither \textsc{2LPTPNG} nor \textsc{FastPM} $k_{\rm max}=1$ correct for aliasing, so we are seeing indeed the aliasing effect in these codes. 

We found a residual oscillatory pattern for \textsc{FastPM} due to differences in the implementation of the interpolation scheme for the input power spectrum. Also, this different interpolation scheme can account for the $10^{-5}$ systematic difference between \textsc{2LPTPNG} and \textsc{MonofonIC} as it propagates via the $\sigma_8$ normalisation of the power spectrum that these codes perform.

An interesting behaviour occurs in \textsc{FastPM} when setting $k_{\rm max}=1/4$ and $k_{\rm max}=2/3$. There is a jump at those frequencies (we show them as vertical lines). These scales correspond to the point where all the PNG signal is removed for smaller scales for these choices of parameters in \textsc{FastPM}. This leads to a nearly constant offset, at a level of $0.1\%$, in the power spectrum due to the differences between the power spectrum in $f_{\rm NL}=0$ and $f_{\rm NL} = 100$. Note that increasing the parameter \textsc{lpt\_nc\_factor} to $2$, which controls the grid size for the LPT calculations, does not modify the pass band filter $k_{\rm max}$.

The code that shows larger variations with respect to \textsc{MonofonIC} is \textsc{LPICOLA}. The larger variations are due to the different white noise generators used in \textsc{LPICOLA} with respect to the reference simulation, as we explained in \autoref{sec:ic_codes}. Nevertheless, the differences are still compatible with cosmic variance and are below $1\%$, even at $k=k_{\rm Nyq}$.

At $z=0$, we found that the non-linear evolution amplified the slight differences we saw in the initial power spectrum. For \textsc{2LPTPNG} and \textsc{FastPM} (in all the configurations), we found an excellent agreement where the differences on the power spectrum are below $0.2\%$ at all scales. \textsc{LPICOLA} shows larger deviations at the level of the $1\%$ for most scales, with only some fluctuations reaching $4\%$ for the least sampled modes due to the differences in the white noise generator (see \autoref{sec:ic_codes}). 

\subsubsection{Bispectrum}
We now analyse the bispectrum results. As in \autoref{sec:results_aliasing}, we focused on the squeezed limit $(k_1\sim k_2 = k;\; k_3=3k_f)$, as most of the local PNG signal resides there, although we also checked other configurations leading to the same conclusions.  The results are shown in  the right panel of \autoref{fig:relerr_combined_codes}. At the initial conditions, we found an excellent agreement between \textsc{2LPTPNG}, \textsc{FastPM} (with $k_{\rm max}=1$ and $k_{\rm max}=2/3$), and \textsc{MonofonIC}, with marginal differences at the level of $0.1\%$ at all scales. The $\sim 3 \%$ variations in the bispectrum of \textsc{LPICOLA} were due to the differences in the white noise generator (see \autoref{sec:ic_codes}).

The largest difference occurs for \textsc{FastPM} with $k_{\rm max}=1/4$. Although the agreement with the other codes and configurations is excellent at large scales, the low-pass filter set to $\phi^2$ precisely at $k=1/4 \; k_{\rm Nyq}$ leads to a $\sim 80\%$ deviation for the bispectrum at smaller scales. This deviation corresponds to switching from the bispectrum for $f_{\rm NL}=100$ (or in general, $f_{\rm NL} \neq 0$) to $f_{\rm NL}=0$,  eliminating the entire PNG signal for smaller scales. This also has dramatic consequences for the clustering of halos, as we discuss below. The same effect would be present for \textsc{FastPM} with $k_{\rm max}=2/3$, but shifted to $k_{cut} = 2/3 k_{nyq}$. However, given our grid resolution and our FFT-based bispectrum estimator (see \autoref{sec:pkb}), we do not see this effect here for $k_{\rm max}=2/3$. 

At $z=0$, we found that the relative differences in the bispectrum for \textsc{FastPM} and \textsc{2LPTPNG} grew to $0.09\%$ yielding again an excellent agreement compared with \textsc{MonofonIC}. Moreover, the strong effect we saw for \textsc{FastPM} $k_{\rm max}=1/4$ was no longer present in the bispectrum at $z=0$. This is because the gravitational evolution part dominates the bispectrum at such late times, so the local PNG bispectrum is subleading. This is not the case at high redshifts, where the bispectrum associated with gravitational evolution is significantly suppressed, and the local PNG bispectrum becomes relevant.

\subsubsection{Halo mass function}

We now focus on how using different initial conditions codes may lead to differences in the clustering of halos. As in \autoref{sec:results_initial_z}, the non-linear nature of gravitational evolution and halo finding amplifies the tiny differences in the initial conditions generated with the different codes, so the resulting halo mass function ratios are very noisy. Although not shown explicitly here, the results are similar to \autoref{fig:hmf_aliasing}. Simulations started with \textsc{MonofonIC}, \textsc{FastPM}, and \textsc{2LPTPNG} agree in the HMF within a $1\%$ for all the mass ranges up to $10^{15} \, h^{-1}M_\odot$, with no systematic deviations. There are fluctuations for \textsc{LPICOLA} for $M_{\rm halo} > 10^{14.85}  \, h^{-1}M_\odot$, reaching the level of $12\%$. However, these differences are not significant as these bins are shot noise dominated, and the cosmic variance due to the different white noise generators can account for this (see \autoref{sec:ic_codes}).

\subsubsection{Halo bias}
To further assess the differences in halo clustering, we split the halos into four mass bins.\footnote{ $1.6\cdot10^{12}  <M(h^{-1}{\rm M}_\odot) <4.0\cdot 10^{12}$,  $4.0\cdot 10^{12}<M(h^{-1}{\rm M}_\odot) <8.0\cdot 10^{12}$, $8.0\cdot 10^{12}<M(h^{-1}{\rm M}_\odot) <2.4\cdot 10^{13}$, and $2.4\cdot 10^{13} <M(h^{-1}{\rm M}_\odot) <1.3\cdot 10^{15}$.} We followed the same strategy as in \autoref{sec:results_aliasing} and opted to compress the information in the halo power spectrum into the linear bias $b_1$ and the PNG-response parameter $p$. 

\begin{figure*}[!htb]
\centering
\includegraphics[width=\textwidth]{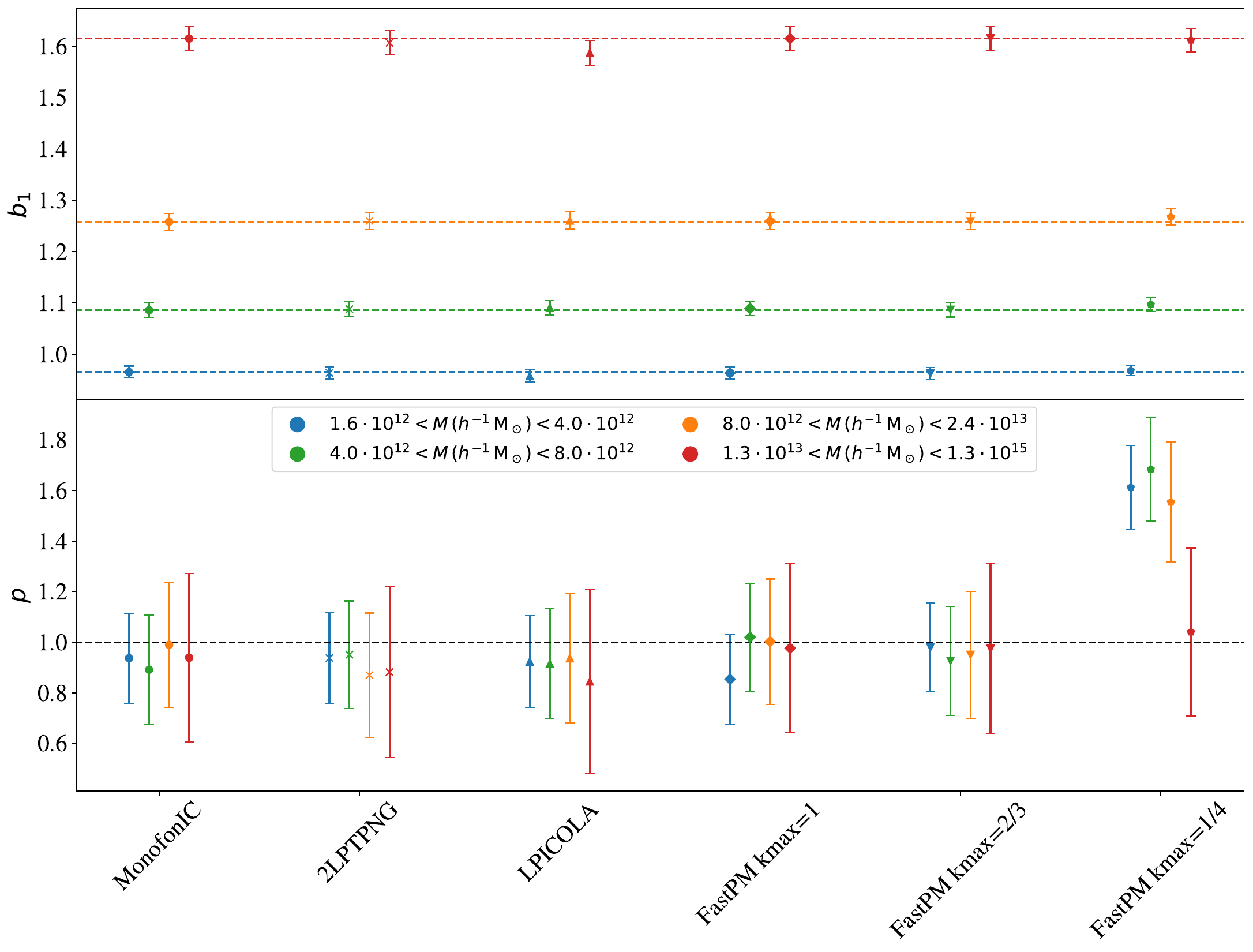}
\caption{Measurements at $z=0$ of the linear bias parameter $b_1$ (upper panel) and the PNG-response parameter $p$ (lower panel) for different codes. All simulations have been initialised at $z=49$ using $2$LPT and $f_{\rm NL} = 100$ (see \autoref{sec:ic_codes}). All the points from the same code or configuration have a different marker and are grouped, from left to right, \textsc{MonofonIC}, \textsc{2LPTPNG},\textsc{LPICOLA}, and \textsc{FastPM}, with $k_{max}=1$, $k_{max}=2/3$, and $k_{max}=1/4$. the colours indicate different halo mass bins (blue: $1.6\cdot10^{12} <M(h^{-1}{\rm M}_\odot) <4.0\cdot 10^{12}$, green: $4.0\cdot 10^{12} <M(h^{-1}{\rm M}_\odot) <8.0\cdot 10^{12} $, orange: $8.0\cdot 10^{12} <M(h^{-1}{\rm M}_\odot) <2.4\cdot 10^{13}$, and red: $2.4\cdot 10^{13} <M(h^{-1}{\rm M}_\odot) <1.3\cdot 10^{15}$). The horizontal dashed lines in the upper panel show the values of $b_1$ measured in our reference simulation, using \textsc{MonofonIC} (\autoref{sec:reference_sims}). In the lower panel, the horizontal dashed line shows the universality relation for $b_\phi/p$. The value of $p$ measured from the simulation run with \textsc{FastPM} assuming $k_{\rm max}=1/4$ is $\sim 2.7\sigma$ away from the reference for the first three mass bins (\autoref{sec:results_aliasing}). The values of $b_1$ and $p$ measured from all other codes and configurations agree within $1\sigma$.}%
\label{fig:b1_p_codes}
\end{figure*}

\autoref{fig:b1_p_codes} shows that all codes produced consistent $b_1$ for all mass bins, compared to the reference simulation (see \autoref{sec:reference_sims}), which uses \textsc{MonofonIC} for generating the initial conditions. Even \textsc{LPICOLA}, with a different initial white noise, shows excellent agreement with the other codes, with only a marginal deviation of $1.1\sigma$  for the most massive bin we considered. This, in turn, is also the one most affected by the shot noise.

Regarding the PNG response parameter $p$, we also found an excellent agreement between \textsc{MonofonIC}, \textsc{2LPTPNG}, \textsc{LPICOLA}, and \textsc{FastPM} (with $k_{\rm max}= 1$ and $2/3$). In particular, we found that for our mass bins, the values of $p$ were compatible within our error bars with the universality relation $p=1$. 

Remarkably, \textsc{FastPM} with $k_{\rm max}=1/4$ showed a tendency to shift $p$ above unity for low-mass halos, indicating a suppression of the scale-dependent bias signal. For high-mass halos, the agreement was again good. We traced this effect to the same scale ($k\approx k_{\rm Nyq}/4$) where the low-pass filter was applied and where we found a $80\%$ deviation in the bispectrum of the initial conditions (see the right panel of \autoref{fig:relerr_combined_codes}). This scale corresponds to a mass of roughly $M\sim 300 \cdot m_{\rm part} \sim 2.4\times 10^{13}\, h^{-1} M_\odot$. In particular, the first three mass bins were below this threshold (for which we recover $p>1$ at $3\sigma$), while for the last mass bin, all halos had a mass above this threshold. 

When the $k_{\rm max}$ parameter was increased to $2/3$ or $1$, this artefact disappeared, as it corresponds to halo masses that are not resolved. We note that although the particular threshold in mass depends on the characteristics of the simulation, the quantity in terms of the mass of the particle is independent of the resolution. We have only analysed the results at $z=0$, so it might be possible that the $M\sim 300 \cdot m_{\rm part}$ threshold varies at higher redshifts.  When using \textsc{FastPM}, our recommendation is to set $k_{\rm max}$ to $2/3$ or $1$, to avoid the spurious effects reported above.

\section{Summary and conclusions}\label{sec:conclusions}

We investigated methods to enhance the accuracy and efficiency of simulations that incorporate local primordial non-Gaussianities (PNG). Here, we focused on: (i) assessing how the initial redshift and LPT order influence the convergence of several statistics in the presence of local PNG; (ii) evaluating aliasing effects due to the local PNG generation; and (iii) comparing the consistency between four state-of-the-art initial condition (IC) generators for PNG, \textsc{MonofonIC} (which served as our reference), \textsc{2LPTPNG}, \textsc{FastPM}, and \textsc{LPICOLA}. The codes are described in \autoref{sec:ic_codes}. 

We produced a suite of $186$ $N$-body simulations with different local PNG amplitudes, $f_{\rm NL} =0,\pm 100$ and $g_{\rm NL} =0,\pm 10^{7}$, in simulation boxes of side $L_{\rm box}= 1\; h^{-1}{\rm Gpc}$, assuming a Planck cosmology (see \autoref{sec:simulations}). We analysed the simulations only at their initial redshift and $z=0$. We ran pairs of aliased and de-aliased simulations, using the $3/2$ zero-padding technique introduced by \cite{Orszag_1971} (\autoref{sec:aliasing}). The simulations have initial redshifts from $z_{\rm start}=99$ down to $z_{\rm start}=11.5$, and LPT orders from $1$ to $3$ (\autoref{tab:simulations}). For a reliable benchmark, we ran a set of reference simulations with eight times more particles, $N_{\rm part} = 1024^3$, than our baseline runs (\autoref{sec:reference_sims}). 

As PNG changes the depth of the gravitational potential wells, the validity of the LPT approximation used to generate the initial conditions needed to be tested.  In \autoref{sec:results_initial_z} we quantified how the initial redshift and LPT order affect the convergence of the power spectrum, bispectrum, halo mass function, halo bias, and PNG response parameter $p$ compared with a high-resolution simulation. Our results for simulations with PNG are fully consistent with those for the Gaussian case, which was already tested in \citet{Michaux_2020}, and are also in line with the findings from \citet{Stahl_2024} for scale-dependent PNG. Pushing to lower $z_{\rm start}$ helped reduce the effect of discreteness errors. 

We find that a configuration with $3$LPT and $z_{\rm start}=11.5$ leads to a sub-percent agreement in the power spectrum up to $k_{\rm Nyq}$ and in the bispectrum up to $2/3 k_{nyq}$. The configuration for the initial conditions has a significant impact on the halo mass function and halo bias. However, this does not affect the PNG response parameter $p$ for halos selected by mass. Our positive and negative $f_{\rm NL}$ and $g_{\rm NL}$ runs show a similar behaviour at $z = 0$ as in the Gaussian case. Thus, given that \citet{Michaux_2020} already showed that discreteness and truncation errors were larger at higher $z$, we expect the PNG simulations to follow the same trend as the Gaussian case at intermediate redshifts.

We quantified the aliasing signal that arises in simulations with local PNG due to the terms $\phi^2$ and $\phi^3$, in \autoref{sec:results_aliasing}. We measured the PNG aliasing signal in the power spectrum, bispectrum, halo mass function, halo bias function, and PNG response parameter by comparing pairs of aliased and de-aliased simulations (\autoref{sec:aliasing}). These pairs have the same configuration beyond the aliasing signal.  
The power spectrum of the initial conditions shows a scale-dependent increase.
This effect is sub-percent at large scales, but becomes $\sim 3\%$ at $k_{\rm Nyq}$ for $g_{\rm NL}=\pm 10^{7}$. We find similar trends for the bispectrum of the initial conditions. However, after evolving at $z=0$, the aliasing signal is suppressed, and it is sub-percent at all scales for all the cosmologies we considered. We do not find any significant variation associated with aliasing for the halo bias, $b_1$, or the PNG response parameter, $p$. Intermediate redshifts are expected to fall between the initial and $z=0$ limits. Moreover, because the aliasing signal scales linearly with the PNG amplitude, the aliasing effect would be even smaller for realistic simulations, including a smaller $|f_{\rm NL}|$ or $|g_{\rm NL}|$ values.

We find sub-percent agreement between the power spectrum and bispectrum (at $z_{\rm start}$ and $z=0$), halo mass function, halo bias, and PNG-response parameters measured in simulations with IC generated with \textsc{MonofonIC}, \textsc{2LPTPNG}, and \textsc{FastPM} with $k_{\rm max}=1$ (\autoref{sec:results_ic_codes}). Given the excellent agreement at both the initial redshift and $z=0$, we expected a similar behaviour at intermediate epochs. After running all the simulations, we found that \textsc{LPICOLA} was producing a different realisation of the white noise when running in the \textsc{GAUSSIAN} mode and when running in \textsc{LOCAL\_FNL} mode. This has complicated the one-to-one comparisons. Nevertheless, the differences found for this code were compatible with cosmic variance. If \textsc{FastPM} is ran with a low-pass filter for the $\phi^2$ field at $k=k_{\rm max}\cdot k_{\rm Nyq}$, our results suggest that a value of $k_{\rm max}\geq 2/3$ should be used to avoid the spurious effects reported in \autoref{sec:results_ic_codes}.

Starting at a lower redshift with higher-order LPT improves the convergence to the fluid limit not only in Gaussian simulations \citep{Michaux_2020} but also for those with PNG. Skipping the first steps of the $N$-body allows more accurate and computationally efficient simulations to be run. This gain is higher if using tree codes, where typically, the first timesteps are the most expensive to run and are also the ones that can introduce more numerical errors during the force evaluations. For example, the run times of the simulations presented here with $z_{\rm start}=99$ were $25\%$ longer than those that started at $z_{\rm start}=11.5$. The limit on pushing for a lower starting redshift must be carefully studied as it depends on the mass resolution.

The aliasing signal induced in simulations with PNG is low. Adequately removing this signal requires adding a zero-padding to the grid in Fourier space, which increases the size of the original grid by a factor of $(3/2)^3$, and thus increases the memory required for the computation. If resources are limited, and $\sim 1\%$ precision is enough, then it is safe to neglect the effect of PNG aliasing. In the future, we will explore the implementation in \textsc{MonofonIC} of an implicit padding algorithm to dealias IC without the extra memory cost \citep[e.g.][]{Bowman_2011}.

We have validated a set of PNG IC generators available to the community. Our results serve to design accurate and efficient simulations with PNG. These will be key for benchmarking analysis tools for the new-generation galaxy surveys, whose goal is to constrain inflationary physics.

\section*{Data availability}

All the codes used in this paper are public. The links to the code repositories are included in footnotes in \autoref{sec:simulations}. All the simulations and data products will be shared under email request to the corresponding author.

\begin{acknowledgements}
      We thank Adrian Bayer for discussions about the draft and helping with FastPM. We also thank Thomas Flöss and Raul Angulo for very useful discussions during the early elaboration of this paper. We also thank the `Centro de Ciencias de Benasque Pedro Pascual' and the `Understanding Cosmological Observations' workshop held within it, as they helped initiating this project. AGA gratefully acknowledges all the members of the Data Science in Astrophysics \& Cosmology Group at the University of Vienna for hosting me as an intern for three months during the development of this work under the Erasmus+Prácticas programme 2023-1-ES01-KA131-HED-000116324.  This work has been supported by Ministerio de Ciencia e Innovaci\' {o}n (MICINN) under the following research grants: PID2021-122603NB-C21 (AGA, VGP, GY) and PID2021-123012NB (AGA, SA). SA has been supported by the Ramon y Cajal fellowship (RYC2022-037311-I) funded by MCIN/AEI/10.13039/501100011033 (Spain) and Social European Funds plus (FSE+). VGP has been supported by the Atracci\'{o}n de Talento Contract no. 2019-T1/TIC-12702 and 2023-5A/TIC-28943 granted by the Comunidad de Madrid in Spain. MM is supported by the MICINN project PID2022-141079NB-C32 The simulations presented in this paper were run in the Hydra cluster of the IFT and in Finisterrae-3 in CESGA under the project AECT-2024-3-0030. The analysis in this work has also been carried out in the computing cluster at UAM (\textsc{taurus}).
\end{acknowledgements}

\bibpunct{(}{)}{;}{a}{}{,} 
\bibliographystyle{aa}
\bibliography{biblio}
\end{document}